\documentclass[journal]{IEEEtran}

\usepackage{cite}
\usepackage{amsmath,amssymb,amsfonts}
\usepackage{algorithmic}
\usepackage{textcomp}
\usepackage{xcolor}
\usepackage{url}
\usepackage[hidelinks]{hyperref}
\usepackage{booktabs}   % 提供 \toprule, \midrule, \bottomrule
\usepackage{multirow}
\usepackage{graphicx}
\usepackage{algorithm}
\usepackage{makecell} 
\usepackage{multirow}
\makeatletter
\let\MYcaption\@makecaption
\makeatother

\usepackage[font=footnotesize]{subcaption}

\makeatletter
\let\@makecaption\MYcaption
\makeatother

\begin{document}

\title{ComHymba: Low-Complexity Domain-Informed Foundation Model for Wireless Communications}

\author{Bowen Yang, 
        Wei Chen, \IEEEmembership{Senior Member, IEEE,}
        Jiaming Cheng,
        Bo Ai, \IEEEmembership{Fellow, IEEE}%
\thanks{Bowen Yang, Wei Chen, Jiaming Cheng, and Bo Ai are with the School of Electronic and Information Engineering, Beijing Jiaotong University, China (email:\{22331171, weich, jiamingcheng, boai\}@bjtu.edu.cn).}}

% make the title area
\maketitle

\begin{abstract}
Wireless foundation models are a promising route to unify channel reconstruction, sensing, and beam management in future wireless communication systems, but existing designs often inherit LLM-style Transformers with quadratic token complexity and weak integration of propagation priors. This paper proposes ComHymba, a domain-informed wireless foundation model built on an asymmetric masked autoencoder for large-scale self-supervised pre-training on Channel State Information (CSI). ComHymba introduces (i) 3D spatio-temporal-frequency patchification with rotary positional embedding, (ii) domain-informed masking strategies that emulate realistic CSI sparsity and fading patterns, and (iii) a decoupled amplitude--phase weighted objective tailored to channel statistics. Architecturally, we employ Hymba blocks that fuse windowed self-attention with state space models (SSMs), enabling linear-time modeling with respect to the overall channel input size. Experiments on eight downstream tasks spanning channel state information reconstruction, environmental sensing, and beam management show consistent accuracy gains over strong task-specific baselines, together with up to a $3.3\times$ inference speedup versus Transformer backbones. Overall, ComHymba provides a scalable and efficient backbone for AI-native physical-layer intelligence.
\end{abstract}


\section{Introduction}

\IEEEPARstart{T}{he} wireless system is entering the sixth-generation (6G) era, where the physical layer is expected to evolve from a pure ``bit-pipe'' into a native platform that jointly supports communications, high-resolution sensing, ubiquitous positioning, and intelligent network operation~\cite{ref:jiang6g,guo2026,chen}. This vision fundamentally enlarges the design space: future networks must handle heterogeneous services and rapidly changing propagation conditions, while processing high-dimensional channel state information (CSI) produced by wideband signals and extremely large antenna arrays. As a result, physical-layer tasks such as channel reconstruction, beam management, environment perception, and mobility-aware optimization become tightly coupled and must be executed under stringent real-time and edge-computing constraints. In this context, building and maintaining isolated, task-specific pipelines is increasingly inefficient, since it duplicates feature extraction, limits cross-task knowledge reuse, and incurs substantial training, storage, and deployment overhead~\cite{big,big2}.

This growing complexity is pushing wireless research toward AI-native solutions. Classical model-based designs rely on simplified analytical assumptions, yet their accuracy degrades in high-frequency, strongly non-stationary, and site-specific channels, leading to pronounced model mismatch~\cite{wuli}. Data-driven deep learning (DL) methods can better capture such non-linearities~\cite{deeple}, but most are still trained in a supervised, per-task manner and thus generalize poorly beyond the training distribution~\cite{fanhua1,fanhua2}. Inspired by the success of large models in other domains, an emerging direction is to pre-train large-scale, self-supervised wireless models that learn transferable representations from massive unlabeled CSI, and then adapt them to diverse downstream tasks with lightweight fine-tuning. Such a ``wireless foundation model'' (WFM) perspective promises a unified intelligence backbone for the physical layer, while simultaneously raising new challenges in architecture efficiency and in embedding radio-domain physical priors~\cite{dingzhi1,dingzhi2}.

Building on this WFM perspective, recent studies have begun to explore large-scale self-supervised pre-training for wireless communications, drawing inspiration from the pre-train--fine-tune paradigm of large language models (LLMs)~\cite{li,st_llm}. By learning from massive unlabeled channel observations, a WFM aims to acquire transferable representations of the wireless environment that can be adapted to heterogeneous downstream tasks with minimal task-specific modification. A key insight is that the wireless channel itself is the shared physical substrate underpinning both communication and sensing~\cite{cfm}. Rather than being a mere collection of numerical entries, CSI serves as a high-fidelity physical carrier that reflects the interaction between electromagnetic waves and the surrounding environment, including geometry, materials, and terminal dynamics. Grounding the foundation model on channel representations therefore encourages the learned features to respect underlying propagation laws, yielding a physically consistent and unified intelligence base that can generalize across frequencies, scenarios, and system configurations.
\begin{figure*}[!t] 
    \centering
    \includegraphics[width=0.9\linewidth]{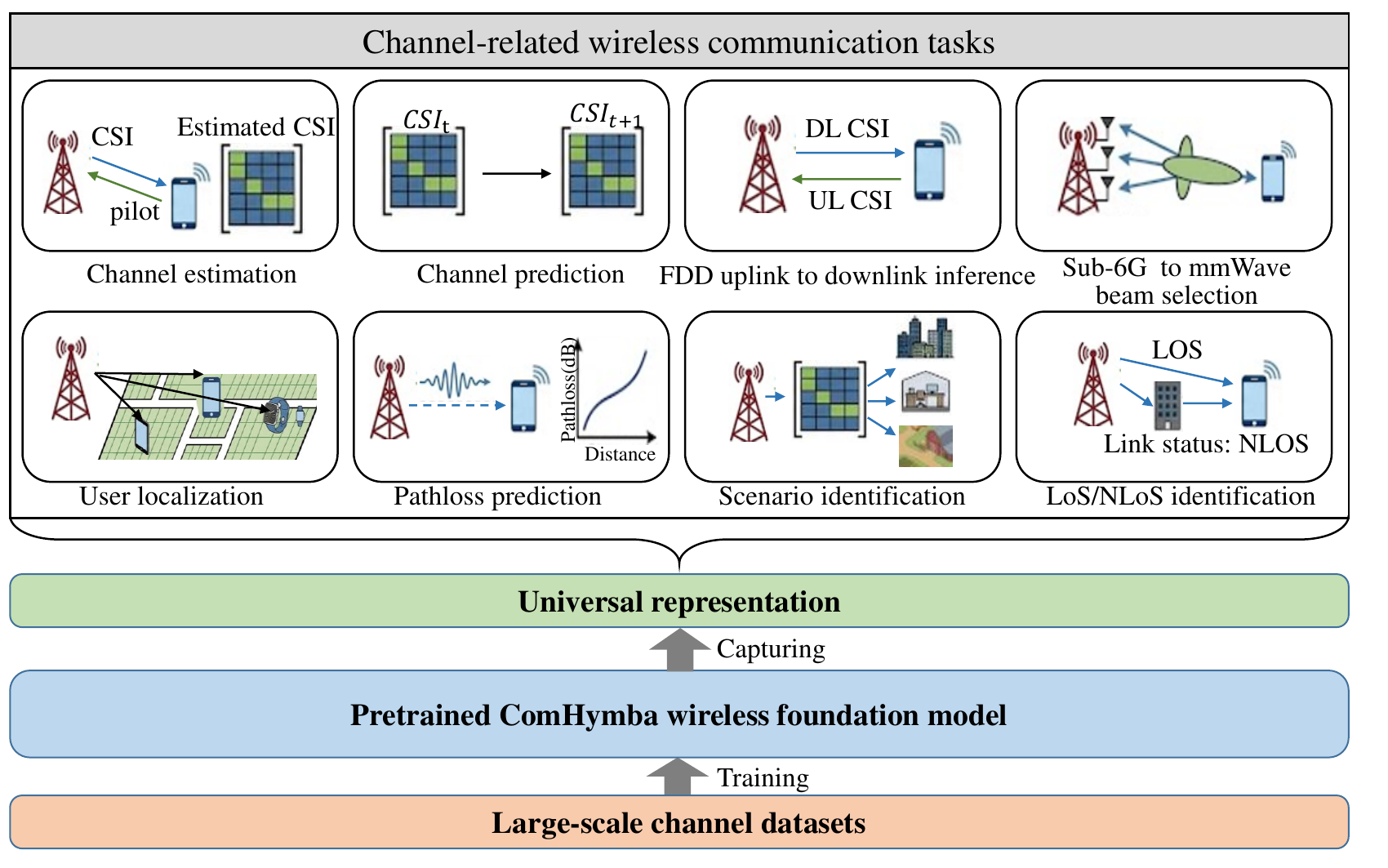}
    \caption{ComHymba for channel-related downstream wireless communication tasks.}
    \label{fig:task}
\end{figure*}

Motivated by the channel-centric view above, a series of early WFMs have been proposed to pursue universal intelligence for the physical layer~\cite{wifo, wirelessgpt,channelgpt,lvmcsi,lwm}. While these attempts validate the promise of large-scale self-supervised pre-training, most existing designs still inherit LLM-style architectures and objectives, without being fully tailored to the computational and domain-specific characteristics of wireless channels. First, their computational footprint often prevents real-time edge deployment. Mainstream WFMs typically adopt standard Transformers, incurring quadratic complexity $O(L^2)$ with the token sequence length $L$~\cite{traneff}; when facing high-dimensional CSI and extremely large antenna arrays, this ``computational wall'' leads to prohibitive training cost and inference latency. Second, the learning criteria are frequently agnostic to radio propagation mechanisms and domain insights. Simply reusing random masking and generic statistical losses makes it difficult to capture the inherent correlations across time, frequency, and space, and it overlooks the distinct roles of amplitude and phase (energy distribution versus structural propagation patterns). Without explicitly embedding such channel-centric domain priors into both architecture and pre-training objective, WFMs are unlikely to deliver high-fidelity and robust physical-layer intelligence across diverse wireless scenarios.

To address the above efficiency and domain-misalignment issues, we propose ComHymba, a low-complexity, domain-informed wireless channel foundation model. ComHymba is built upon a masked autoencoder pre-training framework~\cite{mae} and adopts Hymba blocks \cite{hymba} that couple windowed attention~\cite{attention} with state space models (SSMs)~\cite{ssm}. This hybrid backbone alleviates the Transformer-induced ``computational wall'' by enabling linear-time modeling with respect to the overall channel sequence length, thereby reducing training cost and inference latency for real-time edge deployment. On the learning side, we tailor the entire pre-training pipeline to wireless CSI: (i) a 3D spatio-temporal-frequency patchification and rotary positional encoding to preserve structured correlations, (ii) domain-specific masking schemes (dimension-specific, block-wise, and pilot-pattern) to capture multi-scale cross-dimensional channel correlations, and (iii) a decoupled amplitude--phase weighted objective that combines MSE with energy- and channel-structure-sensitive terms. Together, these designs steer the model to learn representations that are both computationally scalable and domain-consistent, forming a solid foundation for general-purpose physical-layer intelligence.

We validate ComHymba through extensive experiments on eight channel-related downstream tasks spanning three domains: channel reconstruction, environmental sensing, and beam management, as summarized in Fig.~\ref{fig:task}. The results show that ComHymba consistently improves accuracy and robustness over strong task-specific and WFM baselines, while preserving high computational efficiency for practical deployment. The main contributions are summarized as follows:
\begin{itemize}
    \item \textbf{Linear-complexity hybrid backbone:} We propose ComHymba, a low-complexity WFM that combines masked autoencoder pre-training with Hymba blocks, which couple windowed attention and state space models (SSMs). This design reduces the computational complexity from the Transformer's quadratic cost to linear scaling $O(L)$ with sequence length, enabling efficient long-CSI modeling for real-time edge deployment.
    \item \textbf{Domain-informed CSI pre-training pipeline:} We introduce a wireless-tailored pre-training recipe that injects domain priors into representation learning, including 3D spatio-temporal-frequency patchification with rotary positional embedding, domain-specific masking (dimension-specific, block-wise, and pilot-pattern), and a decoupled amplitude--phase weighted objective that augments MSE with energy- and channel-structure-sensitive terms.
    \item \textbf{Broad evaluation across multi-task wireless benchmarks:} We evaluate ComHymba on eight representative downstream tasks covering channel reconstruction, environmental sensing, and beam management. Results show consistent gains in accuracy/robustness with favorable efficiency compared with strong task-specific baselines, supporting ComHymba as a unified backbone for diverse physical-layer tasks.
\end{itemize}

\textit{Notation:} Throughout this paper, we employ the following notation. Boldface italic letters denote tensors, matrices, or vectors. Subscripts are used to indicate indices within specific dimensions, patches, or sets (e.g., $\boldsymbol{H}_i, \Omega_{\mathrm{mask}}$). The superscripts $(\cdot)^T$ and $(\cdot)^H$ represent the transpose and conjugate transpose, respectively, while $\odot$ denotes the Hadamard (element-wise) product. $\mathbb{C}$ and $\mathbb{R}$ signify the complex and real domains, with the imaginary unit denoted by $j = \sqrt{-1}$. We use $|\cdot|$, $\angle(\cdot)$, and $\|\cdot\|_F$ to represent the element-wise absolute value, phase angle, and Frobenius norm, respectively. Square brackets $[\cdot, \cdot]$ are employed for the concatenation of vectors or feature sequences. Additionally, $\mathcal{T}(\cdot)$ and $\Phi$ denote the complex-to-real transformation and the 3D slicing operator, while $O(\cdot)$ characterizes the computational complexity.

\section{System Model and Problem Formulation}

This section presents the system model and learning formulation of ComHymba. We first introduce the underlying wireless channel model and organize CSI into a structured spatio-temporal-frequency tensor representation suitable for large-scale pre-training. We then present a generalized masking--reconstruction objective that casts universal feature learning as an information recovery problem over the multi-dimensional channel index space, providing the mathematical basis for domain-informed self-supervised representation learning.

\subsection{Physical Channel Modeling}
We consider heterogeneous propagation conditions and generate CSI using the standardized geometry-based stochastic model (GBSM) in 3GPP TR~38.901~\cite{3gpp38901}. This provides a physically grounded data source for pre-training.

At a given time instant $t$ and absolute subcarrier frequency $f$, the narrowband frequency-domain multiple-input multiple-output (MIMO) channel is represented by the matrix $\boldsymbol{H}(t,f) \in \mathbb{C}^{N_{\mathrm{rx}}\times N_{\mathrm{tx}}}$, where $N_{\mathrm{tx}}$/$N_{\mathrm{rx}}$ are the transmit/receive antenna counts, and modeled as a superposition of $P$ multipath components:
\begin{equation}
\begin{split}
\boldsymbol{H}(t,f) = & \sum_{p=1}^{P} g_p \, \boldsymbol{a}_{\mathrm{rx}}\left(\theta_p^{\mathrm{rx}}, \phi_p^{\mathrm{rx}}\right) \\
&  \boldsymbol{a}_{\mathrm{tx}}^{H}\left(\theta_p^{\mathrm{tx}}, \phi_p^{\mathrm{tx}}\right) e^{-j 2\pi f \tau_p} e^{j 2\pi \nu_p t},
\end{split}
\end{equation}
where $g_p$ is the complex path gain, $(\theta_p^{\mathrm{tx}},\phi_p^{\mathrm{tx}})$ and $(\theta_p^{\mathrm{rx}},\phi_p^{\mathrm{rx}})$ denote the azimuth/elevation angles of departure and arrival, and $\tau_p$ and $\nu_p$ are the delay and Doppler shift, respectively. The transmit/receive spatial responses are captured by uniform planar array (UPA) steering vectors, which we express via a separable Kronecker structure:
\begin{equation}
\boldsymbol{a}(\theta, \phi) = \boldsymbol{a}_h(\theta, \phi) \otimes \boldsymbol{a}_v(\phi).
\end{equation}
This model makes explicit the coupled structure of CSI across time evolution (Doppler), frequency selectivity (delay), and spatial manifolds (array responses), which ComHymba aims to learn in a unified representation.

\subsection{Multi-Dimensional Tensor Representation}
To expose the inherent structure of CSI for self-supervised pre-training, we stack consecutive channel snapshots over time and frequency into a four-dimensional complex tensor
$\boldsymbol{H} \in \mathbb{C}^{L \times K \times N_{\mathrm{tx}} \times N_{\mathrm{rx}}}$, where $L$ and $K$ denote the numbers of time samples and subcarriers, respectively. This representation explicitly aligns the temporal, spectral, and spatial axes, allowing the model to capture cross-domain correlations (e.g., Doppler-induced temporal dynamics, delay-induced frequency selectivity, and array-manifold structure) in a unified manner.

Since most deep learning operators are implemented in the real domain, we apply a complex-to-real mapping $\mathcal{T}(\cdot)$ and form a real-valued tensor $\boldsymbol{X}$ by concatenating the real and imaginary parts along an additional channel dimension:
\begin{equation}
   \boldsymbol{X} = \mathcal{T}(\boldsymbol{H}) \in \mathbb{R}^{L \times K \times N_s \times 2},
\end{equation}
where $N_s = N_{\mathrm{tx}} \times N_{\mathrm{rx}}$ denotes the aggregated spatial degrees of freedom. This conversion preserves the full complex information of CSI while providing a standardized input format for 3D patchification and masking--reconstruction learning.

\subsection{Multi-Domain CSI Masking and Reconstruction Framework}

We formulate domain-informed representation learning as a self-supervised masking--reconstruction problem over a multi-dimensional CSI index space. Let $\mathcal{I}$ denote the complete index set of the real-valued CSI tensor $\boldsymbol{X} \in \mathbb{R}^{L \times K \times N_s \times 2}$:
\begin{equation}
\begin{aligned}
\mathcal{I} = \{ (t, f, s, c) \mid\;&
1 \le t \le L,\;
1 \le f \le K, \\
&1 \le s \le N_s,\;
c \in \{1, 2\} \}.
\end{aligned}
\end{equation}

During pre-training, a masking operator $\mathcal{M}$ partitions $\mathcal{I}$ into the observed set $\Omega_{\mathrm{obs}}$ and the masked set $\Omega_{\mathrm{mask}}$ (with $\Omega_{\mathrm{obs}} \cup \Omega_{\mathrm{mask}} = \mathcal{I}$ and $\Omega_{\mathrm{obs}} \cap \Omega_{\mathrm{mask}} = \emptyset$). The model input is the partially observed CSI
\begin{equation}
    \boldsymbol{X}_{\mathrm{obs}} = \{ \boldsymbol{x}_i \mid i \in \Omega_{\mathrm{obs}} \},
\end{equation}
and the foundation model $f_{\theta}$ is trained to predict the missing components on $\Omega_{\mathrm{mask}}$ by leveraging structured dependencies in $\Omega_{\mathrm{obs}}$, including temporal dynamics induced by Doppler, frequency correlations induced by multipath delays, and spatial correlations induced by array manifolds.

This masking--reconstruction objective turns pre-training into a domain-consistent inference task: by hiding a subset of spatio-temporal-frequency samples, the model is forced to internalize the joint distribution of CSI across domains, rather than memorizing task-specific input--output mappings. As a result, the learned representation becomes robust to practical imperfections such as sparse pilots, irregular sampling, and partial blockages, providing a principled foundation for universal adaptation to diverse downstream communication-and-sensing tasks.

\begin{figure*}[tbp] % 单栏显示，突出重点
    \centering
    \includegraphics[width=1.0\linewidth]{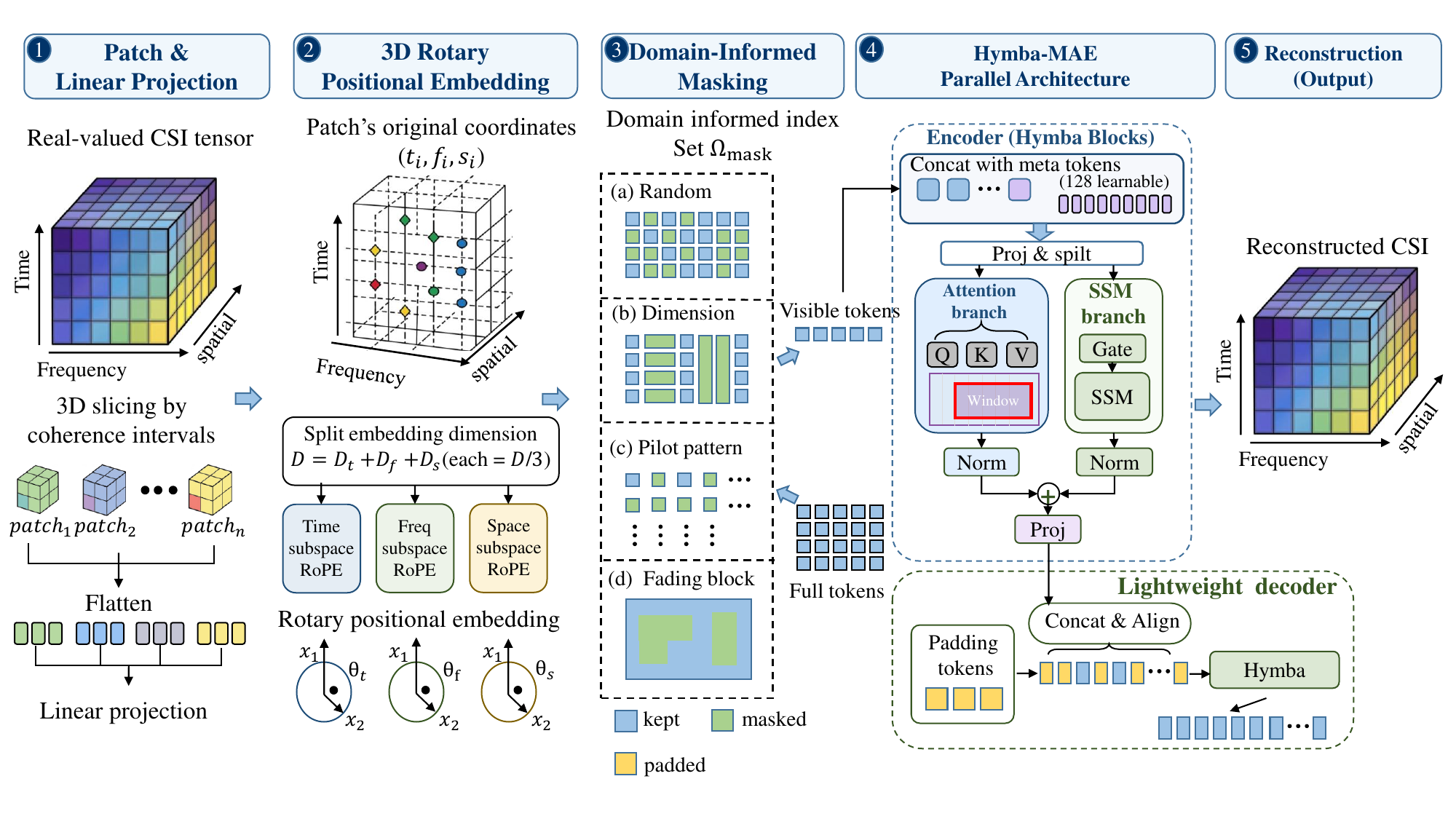}
    \caption{The overall architecture of the proposed ComHymba framework, featuring 3D patchification, domain-informed masking, and the Hymba backbone for wireless channel representation learning.}
    \label{fig:arch}
\end{figure*}
\section{The Proposed ComHymba Framework}

This section presents the ComHymba framework in detail, including both the wireless-tailored pre-training pipeline and the efficient hybrid backbone. As shown in Fig.~\ref{fig:arch}, ComHymba first converts raw CSI into structured spatio-temporal-frequency tokens via 3D patchification, structural embedding, and domain-specific masking, and then performs asymmetric masked reconstruction using a Hymba-based masked autoencoder architecture. By combining reconstruction with linear-complexity sequence modeling, the proposed framework is able to learn domain-consistent channel representations while remaining scalable for long-CSI processing in edge deployments. The key components are described in the following subsections.

\subsection{Spatio-Temporal-Frequency Patchification and Channel Structure Embedding}

Directly ingesting raw CSI samples into a large model is inefficient and often counterproductive. The real-valued CSI tensor $\boldsymbol{X} \in \mathbb{R}^{L \times K \times N_s \times 2}$ is extremely high-dimensional; if each scalar entry is treated as an independent token, the sequence length grows as $L\!\times\!K\!\times\!N_s$, rapidly violating the memory and latency constraints of edge deployment. Meanwhile, such point-wise tokenization amplifies measurement noise and fragments the inherent channel organization (e.g., locally stationary time--frequency regions and spatial coherence), which hinders learning stable spatio-temporal-frequency representations.

To address these issues, we perform spatio-temporal-frequency patchification on the real-valued CSI tensor using a propagation-aware 3D slicing scheme. Let $P_L$, $P_K$, and $P_s$ denote the patch sizes along the time, frequency, and spatial axes, respectively. Since CSI exhibits approximate local stationarity within a small neighborhood in this 3D grid, we define a slicing operator $\Phi$ that partitions $\boldsymbol{X}$ into a set of non-overlapping 3D patches. Compared with point-wise tokenization, this structured partitioning suppresses trivial sample-level perturbations and encourages the model to learn the coherent evolution of local resource blocks, where each patch serves as a compact \emph{channel descriptor} summarizing local propagation characteristics.

This patchification strategy benefits both efficiency and representation learning. From a computational perspective, aggregating $P_L\times P_K\times P_s$ samples into one token shortens the input sequence by the same factor, thereby reducing the cost of modeling long-range dependencies. From a domain perspective, operating at the patch level mitigates sample-wise noise and encourages the model to encode structured propagation patterns (rather than isolated fluctuations). In implementation, each 3D patch is flattened and projected into a $D$-dimensional embedding via a linear layer:
\begin{equation}
    \boldsymbol{z}_i = \text{Linear}(\text{Flatten}(\text{Patch}_{i})) \in \mathbb{R}^D.
\end{equation}

In summary, the proposed spatio-temporal-frequency 3D patchification leverages channel local stationarity to produce a compact token sequence for efficient processing, enabling transferable representation learning with strong domain consistency in pre-training.

\subsection{Spatio-Temporal-Frequency 3D Positional Encoding}
For physical-layer channel modeling, it is crucial to preserve the topological relationships among patches in the original spatio-temporal-frequency grid. However, the masking operation yields a discontinuous observation set, making absolute indexing unreliable and weakening the model’s ability to reason about relative geometry. To remedy this, we introduce a \textit{Spatio-Temporal-Frequency 3D Positional Encoding} based on rotary positional embedding (RoPE). Each patch token is associated with its pre-masking 3D coordinate $(t_i, f_i, s_i)$, enabling the encoder to retain relative displacement cues across time, frequency, and space even under partial observations.

The framework explicitly injects these coordinate positions into the feature vectors through rotational transformations in the complex domain. To achieve decoupled dimensional injection, the $D$-dimensional embedding vector is partitioned into three orthogonal subspaces corresponding to the time, frequency, and space axes, with the constraint that $D$ must be divisible by 6. For each dimension, a rotation angle $\theta$ is generated based on the position index to transform the feature pairs. The frequency base parameter $\omega_k$ is defined as:
\begin{equation}
    \omega_k = 10000^{-\frac{2k}{D/6}}, \quad k \in \{0, 1, \dots, D/6 - 1\}.
\end{equation}

For feature vectors, the rotation process is applied to two-dimensional scalar pairs within each dimensional subspace. For a specific axis, let $pos \in \{t_i, f_i, S_i\}$ denote the original coordinate of the $i$-th patch, and let $x_{1,k}$ and $x_{2,k}$ represent the $k$-th pair of scalar elements in that subspace. The rotation transformation is formulated as follows:
\begin{equation}
    \begin{pmatrix} x'_{1,k} \\ x'_{2,k} \end{pmatrix} = \begin{pmatrix} \cos(pos \cdot \omega_k) & -\sin(pos \cdot \omega_k) \\ \sin(pos \cdot \omega_k) & \cos(pos \cdot \omega_k) \end{pmatrix} \begin{pmatrix} x_{1,k} \\ x_{2,k} \end{pmatrix},
\end{equation}
where $x'_{1,k}$ and $x'_{2,k}$ are the transformed feature components, and the rotation angle is implicitly driven by the patch coordinate $pos$ scaled by the frequency parameter $\omega_k$.

This explicit 3D positional injection leverages the dot-product invariance of RoPE, enabling the model to utilize structural constraints between residual sampling points to preserve channel structural integrity under extreme masking conditions, thereby achieving precise reconstruction of the original channel features.

\subsection{Asymmetric Parallel Hybrid Architecture Design}
This study proposes an asymmetric masked autoencoder architecture based on a parallel hybrid-head structure, utilizing the Hymba block as the core feature extraction unit. Compared with the currently mainstream models that directly stack Transformer operators, the methodological foundation of this architecture lies in the parallel decoupling and synergy of operators at the layer level. This design enables high-fidelity representation of complex radio physical characteristics in the latent space, aiming to significantly enhance computational speed while ensuring channel modeling accuracy. Unlike the single global attention mechanism in conventional Transformer stacking models, the Hymba block integrates attention and state space model heads in parallel within a single layer. This approach achieves linear computational complexity while enabling efficient representation of long-sequence channel data.

\subsubsection{Sequence Feature Embedding and Enhancement}
In the embedding stage, the masked visible sampling sequence $\boldsymbol{X}_\mathrm{vis} \in \mathbb{R}^{L_\mathrm{vis} \times D}$ is first projected into the latent space via a linear mapping layer. To enhance the model's perception of global prior channel knowledge, 128 learnable Meta Tokens $\boldsymbol{R}$ are prepended to the sequence to form the enhanced input sequence $\tilde{\boldsymbol{X}}$. The feature flow process is expressed as:

\begin{equation}
\tilde{\boldsymbol{X}} = \text{Concat}([\boldsymbol{R}, \text{Linear}(\boldsymbol{X}_\mathrm{vis})]).
\end{equation}

The introduction of Meta Tokens serves not only as a carrier for intrinsic propagation priors but also as an attention ``backstop'' in the computational logic. By absorbing redundant weights generated by ``forced attention,'' these tokens effectively mitigate the attention drainage common in mainstream Transformer models, providing a stable global reference anchor for subsequent parallel feature decoupling.

\subsubsection{Parallel Hybrid-Head Encoder Architecture}
Inside the encoder, the enhanced sequence $\tilde{\boldsymbol{X}}$ undergoes parallel splitting in the feature space through a high-dimensional projection layer to achieve explicit operator decoupling:
\begin{equation}
[\boldsymbol{X}_\mathrm{attn}, \boldsymbol{X}_\mathrm{ssm}, \boldsymbol{G}] = \text{Split}(\text{Proj}(\tilde{\boldsymbol{X}})),
\end{equation}
where $\boldsymbol{X}_{\mathrm{attn}}$ is split along the feature dimension to yield the Query ($\boldsymbol{Q}$), Key ($\boldsymbol{K}$), and Value ($\boldsymbol{V}$) vectors, while $\boldsymbol{X}_\mathrm{ssm}$ and the gating signal $\boldsymbol{G}$ flow into the state-space branch. In the attention branch, the output is computed following the scaled dot-product attention mechanism:
\begin{equation}
\boldsymbol{Y}_\mathrm{attn} = \text{Softmax}\left(\frac{\boldsymbol{Q}\boldsymbol{K}^T}{\sqrt{d_k}}\right)\boldsymbol{V}.
\end{equation}

The attention head acts as ``snapshot memory,'' capturing microscopic phase fluctuations in the channel through precise alignment. Simultaneously, the parallel SSM branch utilizes the linear recursive properties of the Mamba-2 operator to model the dynamic evolution of features on the channel manifold. Its transformation flow is represented as:
\begin{equation}
\boldsymbol{Y}_\mathrm{ssm} = \text{Gate}(\boldsymbol{G}) \odot \text{SSM}(\boldsymbol{X}_\mathrm{ssm}).
\end{equation}

The SSM head functions as ``fading memory,'' extracting macroscopic propagation dynamics such as multipath evolution through context summarization. To ensure mathematical stability during the fusion of heterogeneous branches, independent normalization is applied to each output before integration into the latent representation:
\begin{equation}
\boldsymbol{H}_\mathrm{lat} = \text{Proj}(\text{Norm}(\boldsymbol{Y}_\mathrm{attn}) + \text{Norm}(\boldsymbol{Y}_\mathrm{ssm})).
\end{equation}

This parallel structure ensures the encoder can simultaneously encapsulate high-resolution local details and robust long-range contextual information, significantly enhancing feature extraction completeness and processing efficiency.

\subsubsection{Lightweight Decoder and Structured Channel Reconstruction}
In the reconstruction stage, the framework leverages the design advantages of the asymmetric masked autoencoder by constructing a lightweight decoder to reduce redundant overhead. First, the extracted latent variables $\boldsymbol{H}_\mathrm{lat}$ and learnable mask tokens $\boldsymbol{M}$ are concatenated and precisely realigned based on the original 3D coordinate indices:
\begin{equation}
\boldsymbol{H}_\mathrm{full} = \text{Align}(\text{Concat}([\boldsymbol{H}_\mathrm{lat}, \boldsymbol{M}]), \text{Index}_{\mathrm{3D}}).
\end{equation}

This coordinate-aware realignment mechanism ensures the model initiates the reconstruction task at the correct coordinates on the channel manifold. The lightweight decoder then restores missing channel details through the following operation:
\begin{equation}
\hat{\boldsymbol{H}} = \text{Decoder}(\boldsymbol{H}_\mathrm{full}).
\end{equation}

The decoder achieves a highly efficient structured reconstruction of the multi-dimensional wireless channel response, faithfully capturing its underlying radio characteristics.

\begin{figure*}[!t] 
    \centering
    \includegraphics[width=1.0\linewidth]{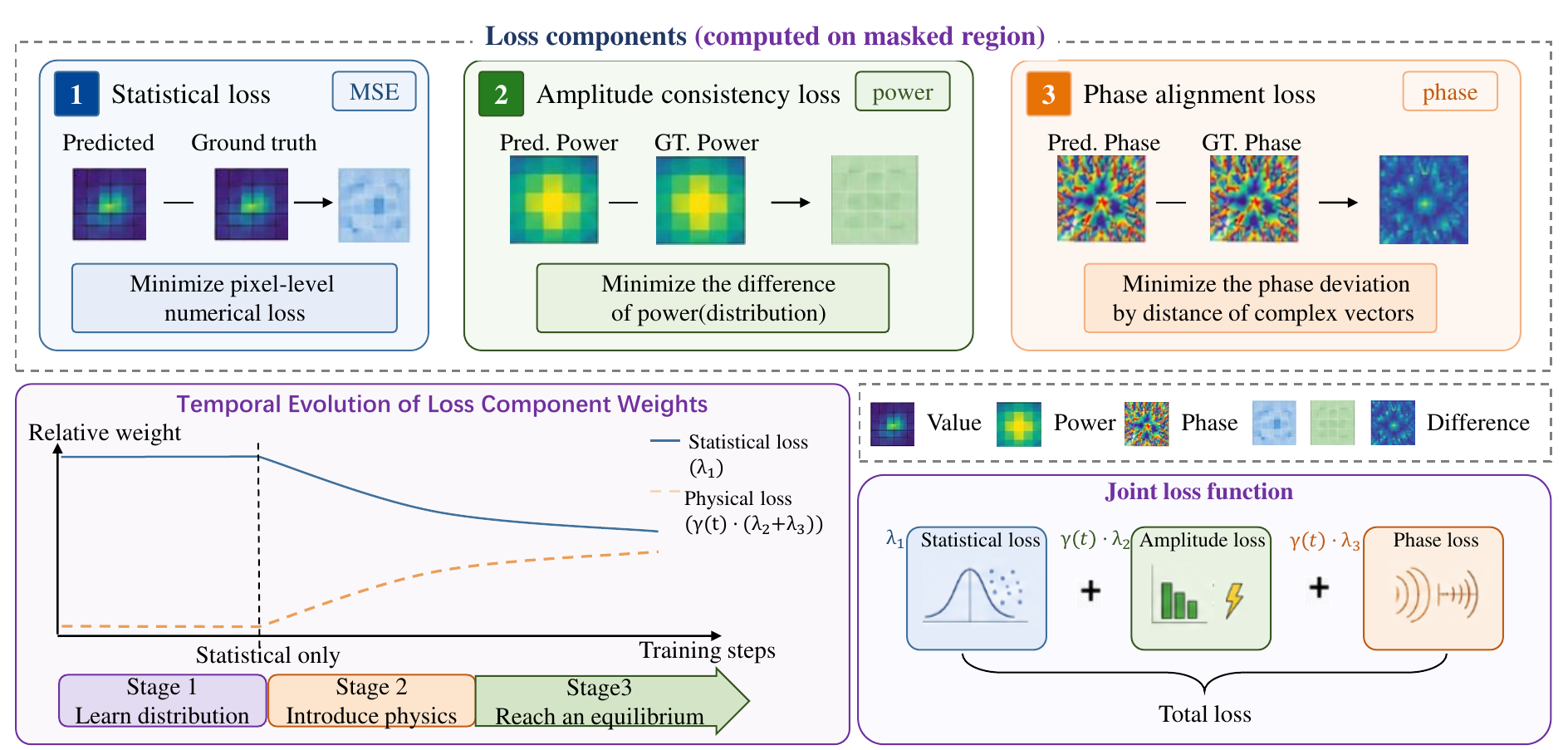}
    \caption{Design of the domain-informed joint loss function.}
    \label{fig:loss}
\end{figure*}

\subsubsection{Analysis of Architectural Advantages: Mechanism of Linear Complexity and Precision Guarantee}
The linearization of computational complexity primarily stems from the restructuring of the mathematical properties of the operators. Traditional self-attention mechanisms exhibit quadratic growth ($O(L^2)$) in overhead relative to the sequence length $L$. In contrast, the proposed architecture employs a sliding window attention mechanism to restrict the search horizon of each point within a fixed width $W$, thereby decoupling the complexity to $O(L)$. Simultaneously, the parallel SSM branch utilizes the linear recursive attributes of the Mamba-2 operator to mathematically ensure that the overall model maintains rapid inference capabilities when processing ultra-high-dimensional data, effectively circumventing computational deficits.

In terms of precision guarantee, the architecture achieves multi-scale representation through the dual-stream synergy of ``snapshot memory'' and ``long-range evolution.'' The attention branch, acting as snapshot memory, captures instantaneous microscopic phase fluctuations and local spatial correlations via a precise alignment mechanism. Meanwhile, the parallel SSM branch serves as the primary engine for capturing long-range context; it utilizes the recurrent dynamic summarization capability of hidden states to transcend local window constraints, extracting macroscopic propagation dynamics such as multipath evolution and environmental variations from long sequences. To further strengthen global stability, Meta Tokens function as ``global anchors'' and ``attention backstops,'' mitigating attention dilution and assisting in the storage of global contextual summaries by providing fixed reference points that do not shift with the sequence.

\subsection{Domain-Specific Masking Strategy: Inducing Electromagnetic Intuition}

Distinct from the unbiased random masking patterns used in general vision or text tasks, the \textit{Domain-Specific Masking} scheme designed in this study is explicitly tailored to conform to the structural characteristics of wireless channels. This scheme moves beyond a purely random token-dropping paradigm by strategically constructing physically-consistent ``blind spots'' in the index space $\Omega_{\mathrm{mask}}$. It forces the model to actively leverage physical gradients and environmental priors at the boundaries of the masked regions during the reconstruction process, thereby achieving a transition from simple numerical completion to deep physical interpolation. The specific strategies are as follows:

\begin{itemize}
    \item \textbf{Multi-scale Random Masking:} Random sampling is performed at high ratios. By destroying the global continuity of the electromagnetic field on a large scale, the model is forced to learn deep joint statistical distributions across the temporal, spectral, and spatial dimensions.
    \item \textbf{Dimension-specific Masking:} Independent masking is applied to specific dimensions, such as continuous temporal slots, spectral sub-bands, or antenna clusters. This strategy aims to cut off the information flow within a particular dimension, compelling the model to mine the inherent evolutionary patterns and physical correlations within that single dimension, thereby strengthening the model's independent representation capability for dimensional features.
    \item \textbf{Pilot-Pattern Masking:} Mimicking the periodic layout of reference signals (e.g., pilots) in communication systems, this strategy executes equidistant comb-type or block-type masking. This forces the model to perform interpolation tasks akin to ``channel estimation'' during the pre-training phase, learning to restore a continuous electromagnetic field from discrete observations on the physical manifold.
    \item \textbf{Deep-Fading Block Masking:} Volumetric masking in 3D space is introduced to simulate real-world obstacles and path loss. By removing contiguous ``space blocks'' to construct \textit{deep-fading} scenarios, the model is compelled to utilize physical gradients from known regions to perform complex field reconstruction.
\end{itemize}

These strategies elevate the pre-training objectives from mechanical data completion to the induction of physical intuition. This ensures that the learned features strictly adhere to electromagnetic propagation laws, laying a solid physical foundation for subsequent multi-task transfer.

\subsection{Domain-Informed Joint Loss Function Design}
The pre-training objective is formulated as a joint loss that couples distribution-level fidelity with radio-physics consistency (Fig.~\ref{fig:loss}). We begin with a \textit{Statistical Loss} $\mathcal{L}_{\mathrm{stat}}$ to let the model learn the underlying channel statistics through masked reconstruction. Specifically, $\mathcal{L}_{\mathrm{stat}}$ is the mean squared error (MSE) computed only over masked patches indexed by $\Omega_{\mathrm{mask}}$:
\begin{equation}
\mathcal{L}_{\mathrm{stat}} = \frac{1}{|\Omega_{\mathrm{mask}}|} \sum_{i \in \Omega_{\mathrm{mask}}} \frac{1}{|\mathcal{E}_i|} \| \boldsymbol{H}_i - \hat{\boldsymbol{H}}_i \|_F^2,
\end{equation}
where $\boldsymbol{H}_i$ and $\hat{\boldsymbol{H}}_i$ are the ground-truth and reconstructed real-valued tensor streams of the $i$-th patch, respectively, $\|\cdot\|_F$ denotes the Frobenius norm, and $|\mathcal{E}_i|$ is the number of sampling points in patch $i$.

As training proceeds, we further inject \textit{Channel Physical Constraint Losses} to explicitly regularize the reconstruction with respect to radio propagation principles, focusing on amplitude--frequency behavior and phase evolution. For this purpose, the reconstructed and reference patch tensors are first converted from their real-valued representations back to the complex domain, where physically meaningful magnitude and phase can be evaluated. The proposed physical constraints include two complementary terms:

\textbf{Amplitude (Energy) Consistency Loss $\mathcal{L}_{\mathrm{eng}}$:} This loss penalizes deviations in the received-power distribution between the reconstructed and ground-truth channels. It is implemented by measuring, via the Frobenius norm, the element-wise difference between the squared magnitudes of the complex-valued patch tensors, thereby encouraging the model to preserve large-scale effects such as path loss and shadowing:
\begin{equation}
\mathcal{L}_{\mathrm{eng}} = \frac{1}{|\Omega_{\mathrm{mask}}|} \sum_{i \in \Omega_{\mathrm{mask}}} \frac{1}{|\mathcal{E}_i|} \| |\boldsymbol{H}_i|^2 - |\hat{\boldsymbol{H}}_i|^2 \|_F^2,
\end{equation}
where the absolute value and squaring operations $|\cdot|^2$ are applied in an element-wise manner over the complex-represented tensors.

\textbf{Phase Alignment Accuracy Loss $\mathcal{L}_{\mathrm{phase}}$:} This loss emphasizes accurate recovery of wavefront geometry and spatial directivity encoded in the channel phase. To mitigate the discontinuity induced by phase wrapping within $[-\pi,\pi]$, we compare phases after mapping them onto the unit circle in the complex plane. The phase deviation is then measured by the Frobenius norm of the difference between the resulting unit-modulus complex tensors:
\begin{equation}
\mathcal{L}_{\mathrm{phase}} = \frac{1}{|\Omega_{\mathrm{mask}}|} \sum_{i \in \Omega_{\mathrm{mask}}} \frac{1}{|\mathcal{E}_i|} \| e^{j \angle \boldsymbol{H}_i} - e^{j \angle \hat{\boldsymbol{H}}_i} \|_F^2,
\end{equation}
where $\angle(\cdot)$ extracts the element-wise phase and $e^{j(\cdot)}$ performs the corresponding unit-modulus complex embedding for each patch tensor.

To improve optimization stability, we adopt a progressive strategy that introduces the physical constraints only after the model has learned a reliable statistical prior. Concretely, once training reaches a predefined plateau, the physical losses are activated through a scheduling factor $\gamma(t)$ and merged into the overall objective. The resulting joint loss is
\begin{equation}
\mathcal{L}_{\mathrm{total}} = \lambda_1 \mathcal{L}_{\mathrm{stat}} + \gamma(t) \left( \lambda_2 \mathcal{L}_{\mathrm{eng}} + \lambda_3 \mathcal{L}_{\mathrm{phase}} \right),
\end{equation}
where $\gamma(t)$ increases over time to gradually emphasize propagation-aware regularization, and $\lambda_1, \lambda_2, \lambda_3$ are hyperparameters that balance the three terms. This dual-guided objective helps \textit{ComHymba} learn representations that are both statistically faithful and physically plausible, thereby benefiting downstream tasks.

To clarify the execution of the entire training workflow, the detailed pre-training procedure of the proposed ComHymba framework---integrating data preprocessing, domain-informed masking, hybrid feature extraction, and physics-consistent reconstruction---is formally summarized in \text{Algorithm}~\ref{alg:ComHymba_framework}.

\begin{algorithm}[t]
\caption{The ComHymba Framework Pre-training Procedure}
\label{alg:ComHymba_framework}
\begin{algorithmic}[1]
\renewcommand{\algorithmicrequire}{\hspace*{0.45cm} \textbf{Input:}}
\renewcommand{\algorithmicensure}{\hspace*{0.45cm} \textbf{Output:}}

\REQUIRE Raw complex CSI tensor $\boldsymbol{H} \in \mathbb{C}^{L \times K \times N_{\mathrm{tx}} \times N_{\mathrm{rx}}}$.
\ENSURE Pre-trained ComHymba foundation model weights $\theta$.

\STATE \textbf{Stage I: Data Preprocessing}
\STATE Convert $\boldsymbol{H}$ to real-valued tensor $\boldsymbol{X}$ via transformation $\mathcal{T}(\cdot)$.
\STATE Perform 3D slicing $\Phi$ to generate full 3D patches $\{Patch_i\}$.
\STATE Map all patches to $D$-dimensional embeddings $\boldsymbol{z}_i$ via Linear Projection.
\STATE Inject 3D coordinates $(t_i, f_i, s_i)$ into $\boldsymbol{z}_i$ via 3D RoPE.

\STATE \textbf{Stage II: Domain-Informed Masking}
\STATE Select a domain-specific masking strategy (e.g., Pilot-pattern, Deep-fading).
\STATE Generate masked index set $\Omega_{\mathrm{mask}}$ and visible set $\Omega_{\mathrm{obs}}$.
\STATE Retain only visible patches $\boldsymbol{z}_{\mathrm{vis}} = \{\boldsymbol{z}_i \mid i \in \Omega_{\mathrm{obs}}\}$ as input sequence.

\FOR{each training step}
\STATE \textbf{Stage III: Hybrid-Head Encoder}
    \STATE Prepend Meta Tokens $R$ to form enhanced sequence $\tilde{\boldsymbol{X}} = [\boldsymbol{R}, \boldsymbol{z}_{\mathrm{vis}}]$.
    \STATE $[\boldsymbol{X}_{\mathrm{attn}}, \boldsymbol{X}_{\mathrm{ssm}}, \boldsymbol{G}] = \mathrm{Split}(\mathrm{Proj}(\tilde{\boldsymbol{X}}))$.
    \STATE \textbf{Branch A (Attention):} Compute $\boldsymbol{Y}_{\mathrm{attn}}$ via windowed self-attention to capture snapshot phase fluctuations.
    \STATE \textbf{Branch B (SSM):} Compute $\boldsymbol{Y}_{\mathrm{ssm}}$ via Mamba-2 recursive operator to capture long-range fading evolution.
    \STATE \textbf{Gating \& Fusion:} $\boldsymbol{H}_{\mathrm{lat}} = \mathrm{Proj}(\mathrm{Norm}(\boldsymbol{Y}_{\mathrm{attn}}) + \mathrm{Norm}(\boldsymbol{Y}_{\mathrm{ssm}}))$.
\STATE \textbf{Stage IV: Consistency Reconstruction \& Optimization}
\STATE Realign $\boldsymbol{H}_{\mathrm{lat}}$ with learnable mask tokens $\boldsymbol{M}$ based on original 3D indices.
\STATE Reconstruct complete channel state matrix $\hat{\boldsymbol{H}}$ using lightweight decoder.
\STATE Update $\theta$ by minimizing $\mathcal{L}_{\mathrm{total}}$ (Joint Statistical and Physical Losses).
\ENDFOR
% \RETURN $\theta$
\end{algorithmic}
\end{algorithm}

\section{Experiments and Results Analysis}

This section presents the experimental setup used to assess the representation learning capability and transferability of the proposed \textit{ComHymba} foundation model. The evaluation pipeline is organized to mirror practical deployment, progressing from dataset construction and pre-training to downstream adaptation. Specifically, we first describe the heterogeneous multi-scenario CSI corpus and the corresponding simulation/parameter settings. We then introduce the full pre-training recipe, including the optimization details and the joint statistical--physical objective. Finally, we transfer the pre-trained encoder to eight representative wireless downstream tasks spanning channel reconstruction, environmental sensing, and beamforming. All results are reported with quantitative comparisons against task-specific baselines to demonstrate that \textit{ComHymba} supports heterogeneous tasks within a unified architecture while delivering consistent performance gains.

\subsection{Dataset Configuration}
\begin{figure}[!t] 
    \centering
    \includegraphics[width=1.0\linewidth]{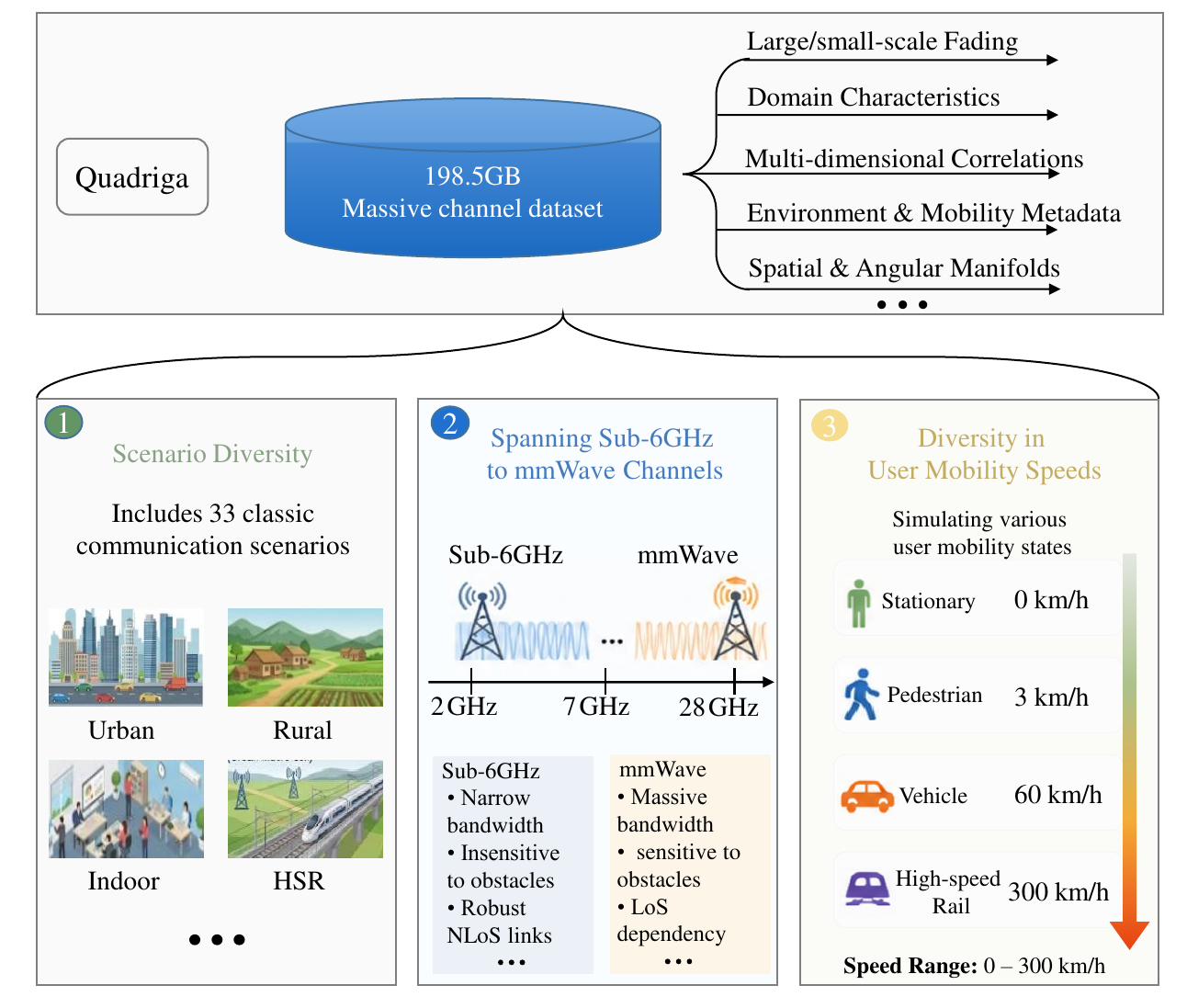}
    \caption{Overview of the massive heterogeneous wireless channel dataset. The dataset comprises 198.5 GB of raw CSI data across 33 diverse scenarios, spanning a wide frequency range from sub-6GHz to mmWave and covering multi-level user mobility speeds from 0 to 300 km/h.}
    \label{fig:data}
\end{figure}
To evaluate the self-supervised pre-training effectiveness of \textit{ComHymba}, we build a large-scale heterogeneous CSI corpus of 198.5~GB using the QuaDRiGa simulator~\cite{quadriga} (Fig.~\ref{fig:data}). The corpus covers 33 representative propagation scenarios and follows 3GPP-compliant configurations to ensure physically grounded and reproducible channel statistics. The generated CSI preserves key propagation characteristics, including multi-path fading, delay profiles, and spatial correlation across antenna elements. To emulate realistic non-stationary environments, we sweep a wide range of carrier frequencies (from sub-6~GHz to mmWave), antenna array geometries, and user mobility patterns/speeds, yielding channels with diverse spatiotemporal dynamics. This heterogeneous corpus provides a strong basis for learning scenario-invariant representations that can be transferred to downstream tasks in channel reconstruction, environmental sensing, and beamforming.

\subsection{Training Configuration and Masking Rate Schedule}

To stably pre-train \textit{ComHymba} on the 198.5~GB heterogeneous CSI corpus, we adopt a carefully tuned optimization protocol whose key hyperparameters are summarized in Table~\ref{tab:training_params}. We use AdamW optimizer with a hybrid learning rate schedule, which initiates from a base learning rate of $1 \times 10^{-5}$ and undergoes linear warmup over the first 5\% of the total training steps up to a peak value of $1 \times 10^{-4}$, followed by cosine annealing to $1 \times 10^{-6}$. Automatic mixed precision (AMP) is enabled throughout to reduce memory cost and accelerate convergence while preserving numerical stability for complex-valued channel representations.

We further employ a curriculum-style masking schedule to gradually increase task difficulty. The masking ratios for multi-scale random masking, dimension-specific striping, and deep-fading block masking are linearly ramped from 50\% to 75\% across training, encouraging the model to first capture coarse time--frequency--spatial regularities and then learn more challenging cross-dimensional propagation dynamics under severe information sparsity. In contrast, pilot-pattern masking is kept fixed at a high-sparsity setting, retaining only one visible element within each local $2 \times 2 \times 2$ cube as a reference signal along the time, frequency, and spatial axes.
\begin{table}[!t]
\caption{ComHymba Pre-training Hyperparameters.}
\label{tab:training_params}
\centering
\begin{tabular}{lc}
\toprule
\textbf{Hyperparameter} & \textbf{Value} \\
\midrule
Batch Size              & 128 \\
Total Epochs            & 250 \\
Optimizer               & AdamW \\
Base Learning Rate           & $1 \times 10^{-5}$ \\
Max Learning Rate      & $1 \times 10^{-4}$ \\
Min Learning Rate       & $1 \times 10^{-6}$ \\
LR Schedule             & Linear Warmup + Cosine Annealing \\
Warmup Ratio            & 5\% of total steps \\
\bottomrule
\end{tabular}
\end{table}

\subsection{Downstream Task Adaptation and Performance Evaluation}

To quantitatively evaluate the transferability and generalization of ComHymba, we apply it to eight core downstream tasks across three categories: channel reconstruction, environmental sensing, and beamforming. By attaching lightweight task-specific heads to the pre-trained encoder, we analyze the model's effectiveness in handling diverse wireless physical layer dimensions.

\subsubsection{Model Configuration and Comparison Baselines}

ComHymba is configured with approximately 100M parameters, featuring an asymmetric architecture with a 20-layer encoder and a 4-layer decoder ($D=504$). To balance performance and overhead, it employs a hybrid attention mechanism: full attention is used in critical layers for global anchoring, while the remaining layers utilize lightweight local window attention for feature evolution.

To comprehensively measure performance, the following representative models are introduced as baselines, including a foundation model of comparable scale:

\begin{itemize}
    \item \textbf{Codebook:} A pre-defined beam vector set based on standards, serving as a training-free physical performance benchmark for beamforming.
    \item \textbf{Multi-Layer Perceptron (MLP):} Simple and low-latency, but lacks perception of structural correlations. Used for FDD inference, path loss prediction, user localization, LoS/NLoS identification, and beamforming~\cite{mlp-beam,mlp-ds}.
    \item \textbf{Convolutional Neural Network (CNN):} Excels at capturing local spatial and time-frequency features via parameter sharing. Used for FDD inference, user localization, path loss prediction, LoS/NLoS identification, and scenario classification~\cite{cnn-fd,cnn-ds}.
    \item \textbf{RNN (LSTM/GRU):} Specialized in modeling time-series characteristics and non-linear dynamics. Used for channel prediction, channel estimation~\cite{lstm,gru}.
    \item \textbf{Transformer \& WIT:} Built on self-attention for global dependency modeling; WIT is specifically optimized for sensing. Used for channel prediction, user localization, and path loss prediction~\cite{wit,transformer_pr}.
    \item \textbf{LLM4CP:} Serving as a foundation model baseline of comparable scale to ComHymba, this architecture is optimized for communication sequences with high transfer potential. Used for channel prediction, channel estimation, FDD inference, scenario classification, and beamforming~\cite{llm4cp,llm4wm}.
    \item \textbf{Helena:} A deep learning architecture tailored for channel estimation, utilizing layer-wise optimizations for physical structure reconstruction~\cite{helena}.
\end{itemize}

\subsubsection{Downstream Task Evaluation and Performance Analysis}
To validate the feature migration and generalization capabilities of the \text{ComHymba} foundation model, this section evaluates its performance across various downstream tasks. The evaluation logic follows a hierarchical progression from low-level signal restoration (micro-level) to environmental semantic extraction (mid-level) and finally to transmission decision control (macro-level). These comparative experiments aim to quantify the core advantages of \text{ComHymba}'s universal representations over raw observation data in the feature space.

To ensure a rigorous quantitative evaluation, we define three key performance metrics across the downstream benchmarks:
\begin{itemize}
    \item \textbf{Normalized Mean Squared Error (NMSE):} Employed as the core evaluation metric for continuous channel reconstruction tasks to quantify the restoration fidelity. The mathematical formulation of the NMSE is defined as:
    \begin{equation}
    \text{NMSE} = \frac{\|\boldsymbol{H}_{\text{gt}} - \boldsymbol{H}_{\text{re}}\|_F^2}{\|\boldsymbol{H}_{\text{gt}}\|_F^2},
    \end{equation}
where $\boldsymbol{H}_{\text{gt}}$ and $\boldsymbol{H}_{\text{re}}$ denote the ground-truth complex CSI tensor and the model's reconstructed output tensor, respectively, and $\|\cdot\|_F$ represents the Frobenius norm.

    \item \textbf{Mean Absolute Error (MAE):} Used for environmental sensing spatial regression tasks, expressed as:
    \begin{equation}
        \text{MAE} = \frac{1}{N_{\text{test}}} \sum_{n=1}^{N_{\text{test}}} |y_n - \hat{y}_n|,
    \end{equation}
    where $y_n$ and $\hat{y}_n$ represent the target and predicted physical parameters for the $n$-th test sample, and $N_{\text{test}}$ denotes the total number of samples in the test set.

    \item \textbf{Classification Accuracy:} Adopted for link-state identification, scenario classification, and top-$K$ beam management choices, formalized as:
    \begin{equation}
        \text{Accuracy} = \frac{1}{N_{\text{test}}} \sum_{n=1}^{N_{\text{test}}} \mathbb{I}(\hat{c}_n \in \mathcal{C}_n^{\text{target}}),
    \end{equation}
    where $\hat{c}_n$ is the model's predicted class or beam index, $\mathcal{C}_n^{\text{target}}$ is the set of legitimate labels (or the top-$K$ optimal beam set), and $\mathbb{I}(\cdot)$ is the indicator function that outputs $1$ if the condition is true and $0$ otherwise.
\end{itemize}

The specific configurations and corresponding performance comparisons of the three task domains are thoroughly presented as follows.

\begin{figure*}[!t]
    \centering
    % 子图 (a): 预测
    \begin{subfigure}[b]{0.38\textwidth}
        \centering
        \includegraphics[width=\linewidth]{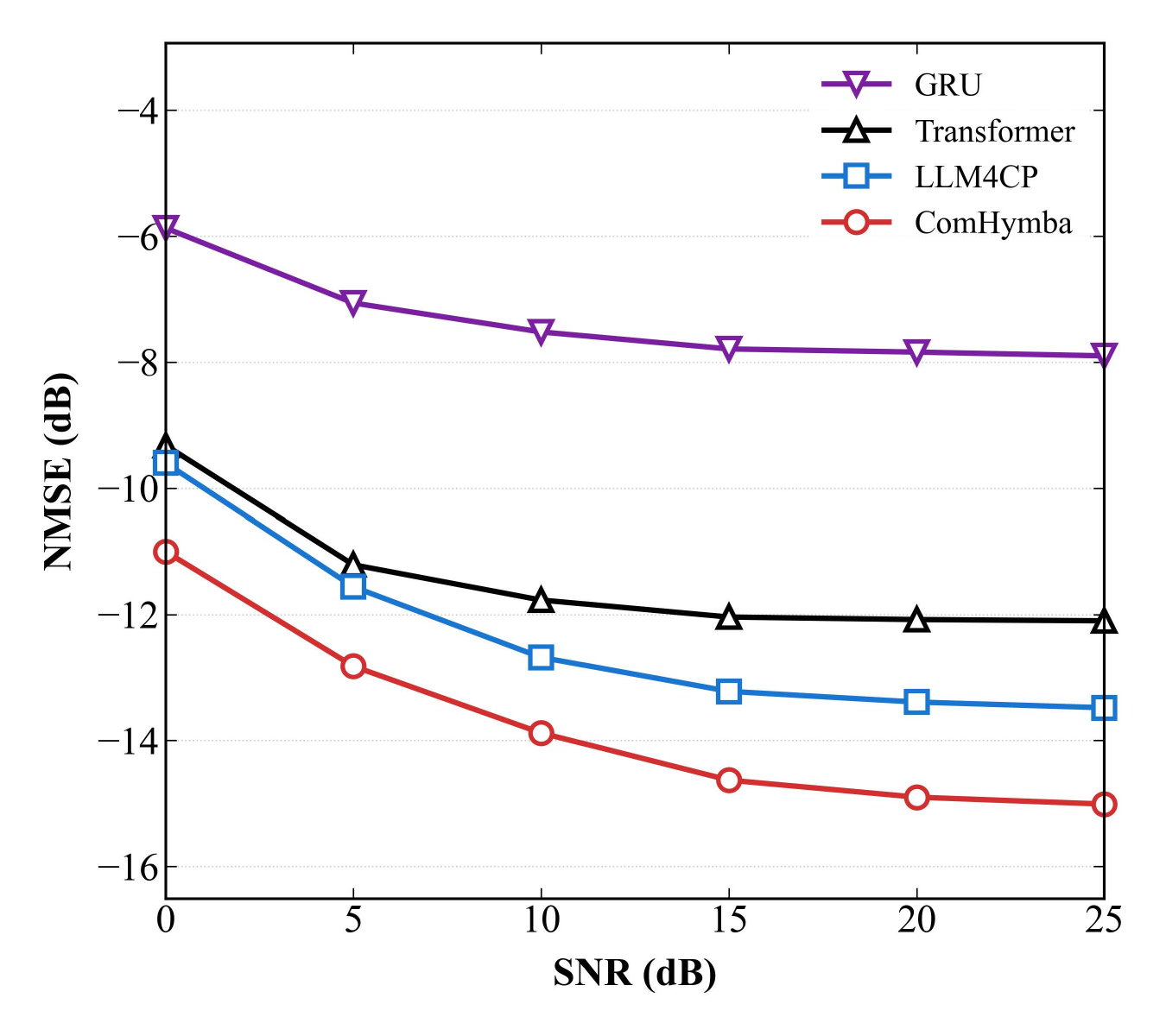}
        \caption{Channel prediction performance under varying SNR levels.}
        \label{fig:pred_nmse}
    \end{subfigure}
    \hspace{0.08\linewidth}
    % 子图 (b): 估计
    \begin{subfigure}[b]{0.38\textwidth}
        \centering
        \includegraphics[width=\linewidth]{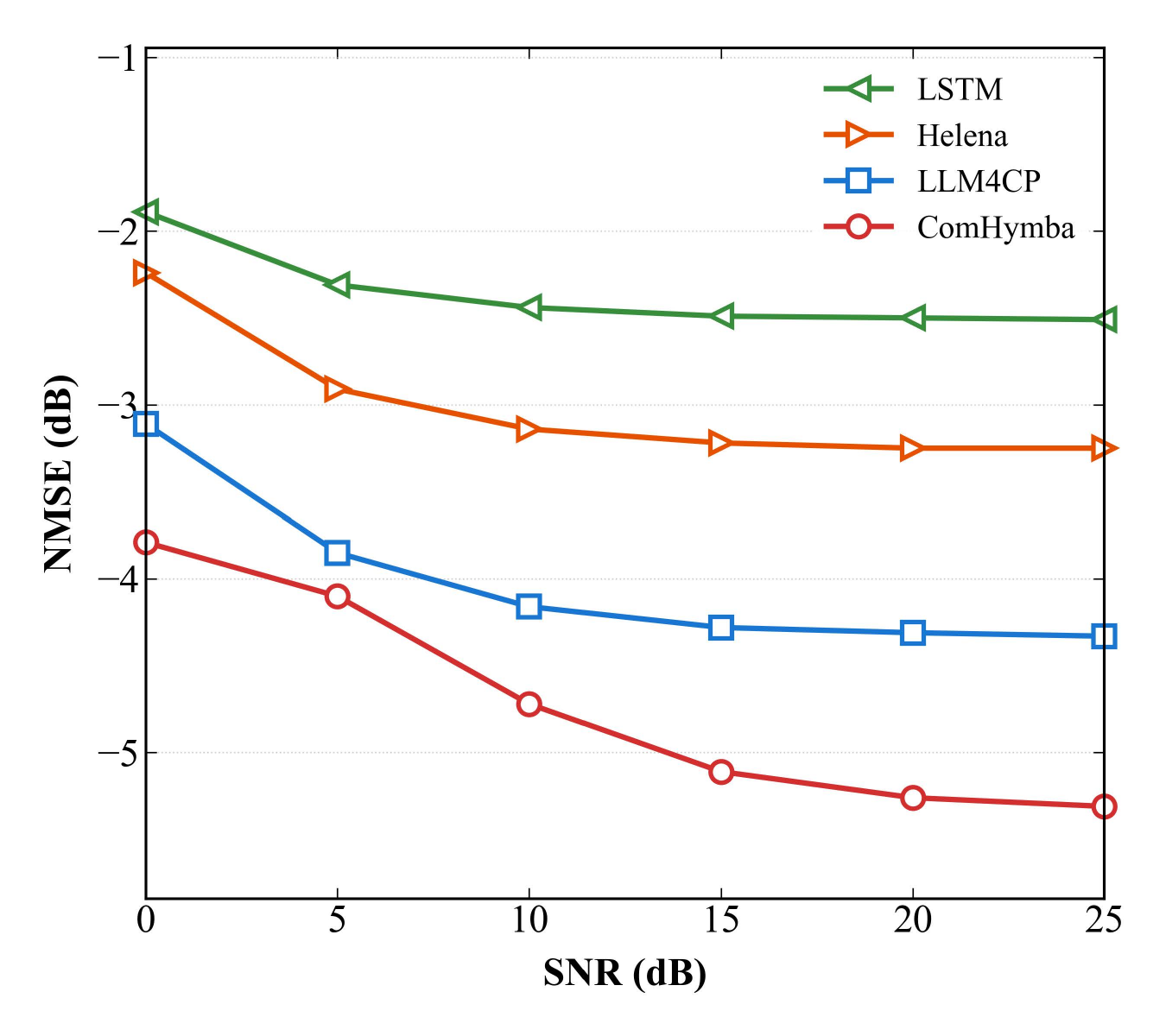}
        \caption{Channel estimation performance under varying SNR levels.}
        \label{fig:est_nmse}
    \end{subfigure}
    \caption{NMSE performance versus SNR for in-band reconstruction tasks: (a) channel prediction and (b) channel estimation.}
    \label{fig:group_inband}
\end{figure*}

\begin{figure}[!t] % 单栏显示，突出重点
    \centering
    \includegraphics[width=0.76\linewidth]{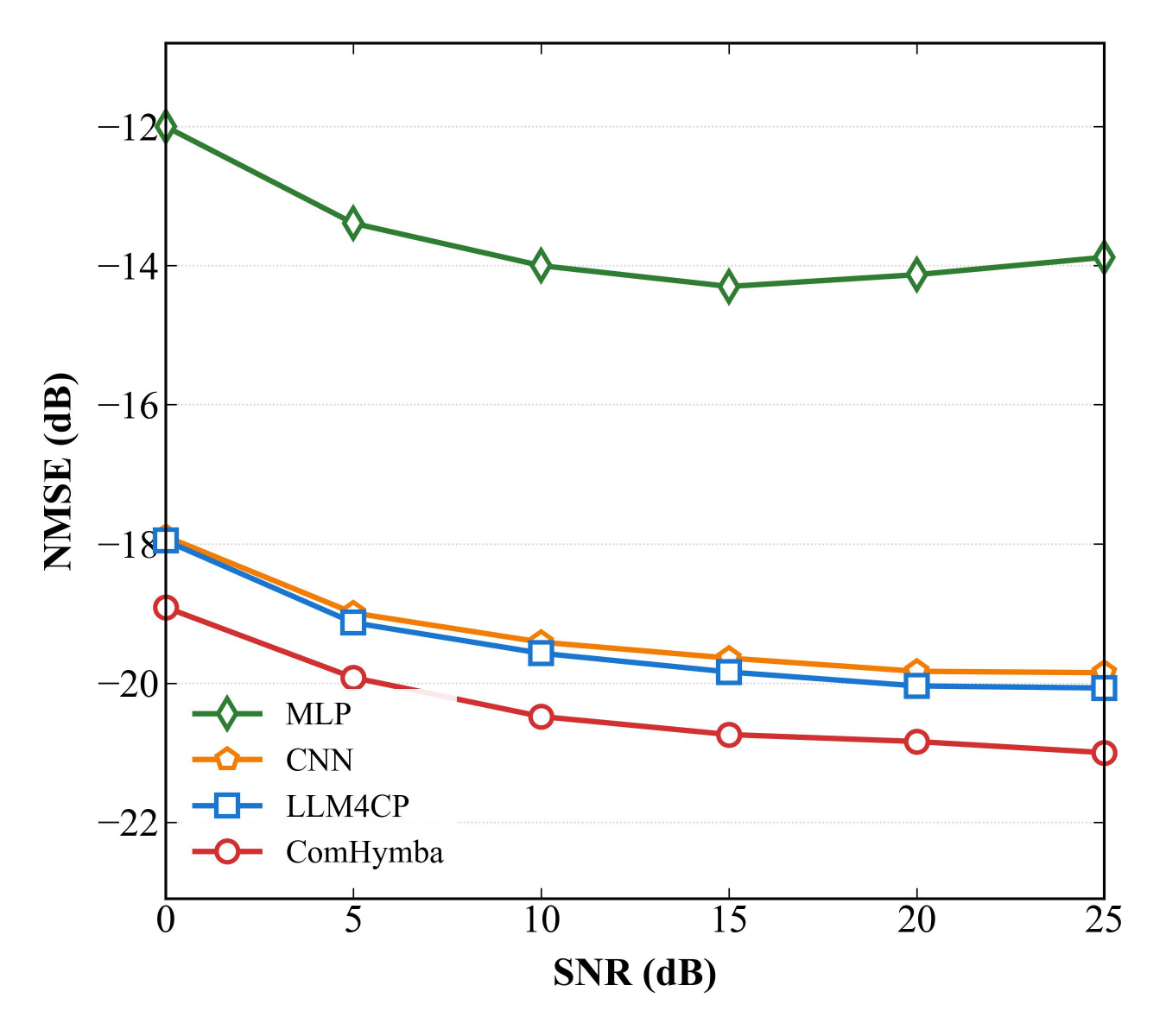}
    \caption{FDD Uplink-to-Downlink Inference: NMSE performance for cross-band mapping.}
    \label{fig:fdd}
\end{figure}

\paragraph{Channel Reconstruction Tasks}

The channel reconstruction tasks aim to verify the capability of the ComHymba foundation model in restoring the underlying physical mechanisms of wireless signals. To ensure the rigor of the evaluation, all sub-tasks in this section are constructed based on a unified OFDM system configuration: the center frequency is set to $2.0 \text{ GHz}$, with a transmission bandwidth of $10 \text{ MHz}$ and a subcarrier spacing of $15 \text{ kHz}$. Regarding the antenna architecture, the base station is equipped with 32 physical antenna elements mapped to 8 logical ports via a specific coupling matrix, while the mobile terminal features 4 receiving antennas, thereby forming a complex $8 \times 4$ MIMO spatial observation model. Based on this system framework, we designed three core sub-tasks.
\begin{itemize}
    \item The channel prediction task is designed to address the "channel aging" problem. It requires the foundation model to decouple stable multipath evolution features from the universal representations generated by ComHymba and suppress cumulative prediction errors within extremely short coherence times, providing real-time link pre-compensation for the system.

    \item The channel estimation task is the cornerstone of coherent demodulation. This task requires the model to utilize the time-frequency physical correlation priors learned during the pre-training phase to achieve precise completion of the fine manifold structure of the channel through the "super-resolution" reconstruction capability of its internal representations, thereby significantly releasing spectrum resources occupied by pilots.

    \item The FDD inference task is dedicated to breaking the channel reciprocity mismatch. This inference is intended to verify ComHymba's ability to extract universal features, thereby reconstructing downlink CSI without the need for downlink feedback to support efficient FDD downlink precoding.
\end{itemize}

In channel reconstruction tasks, ComHymba demonstrates exceptional representation capabilities. For channel prediction (Fig. \ref{fig:pred_nmse}), even under multi-step inference, ComHymba achieves a $10\%\text{--}15\%$ accuracy improvement across the entire SNR range compared to LLM4CP, reflecting its robustness in capturing non-linear dynamics. In channel estimation (Fig. \ref{fig:est_nmse}), under an extremely sparse $2.3\%$ pilot density, its accuracy achieves a $60\%$ leap over the task-specific Helena baseline, proving its power to transform cluttered raw samples into highly structured representations. Furthermore, in the FDD inference task (Fig. \ref{fig:fdd}), facing asymmetric high-dimensional mapping, ComHymba delivers a performance gain of over $15\%$ compared to traditional models, validating its generalization ability to extract frequency-independent physical essences for precise cross-band inference.
\paragraph{Environmental Sensing Tasks}

\begin{figure*}[t]
    \centering
    % 第一行
    \begin{subfigure}[b]{0.38\textwidth}
        \centering
        \includegraphics[width=\linewidth]{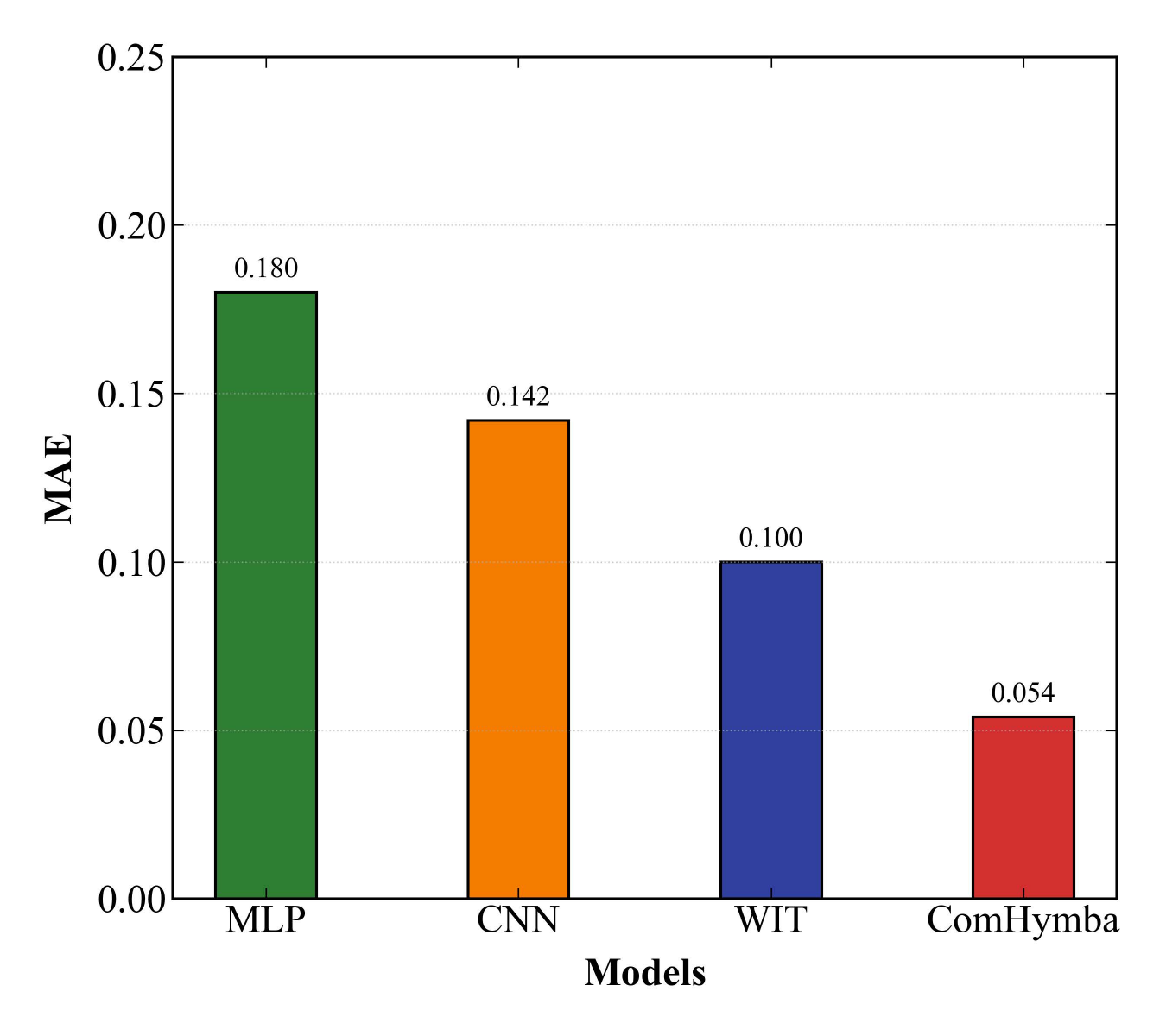}
        \caption{MAE for 3D user localization.}
        \label{fig:loc_mae}
    \end{subfigure}
    \hspace{0.08\linewidth}
    \begin{subfigure}[b]{0.38\textwidth}
        \centering
        \includegraphics[width=\linewidth]{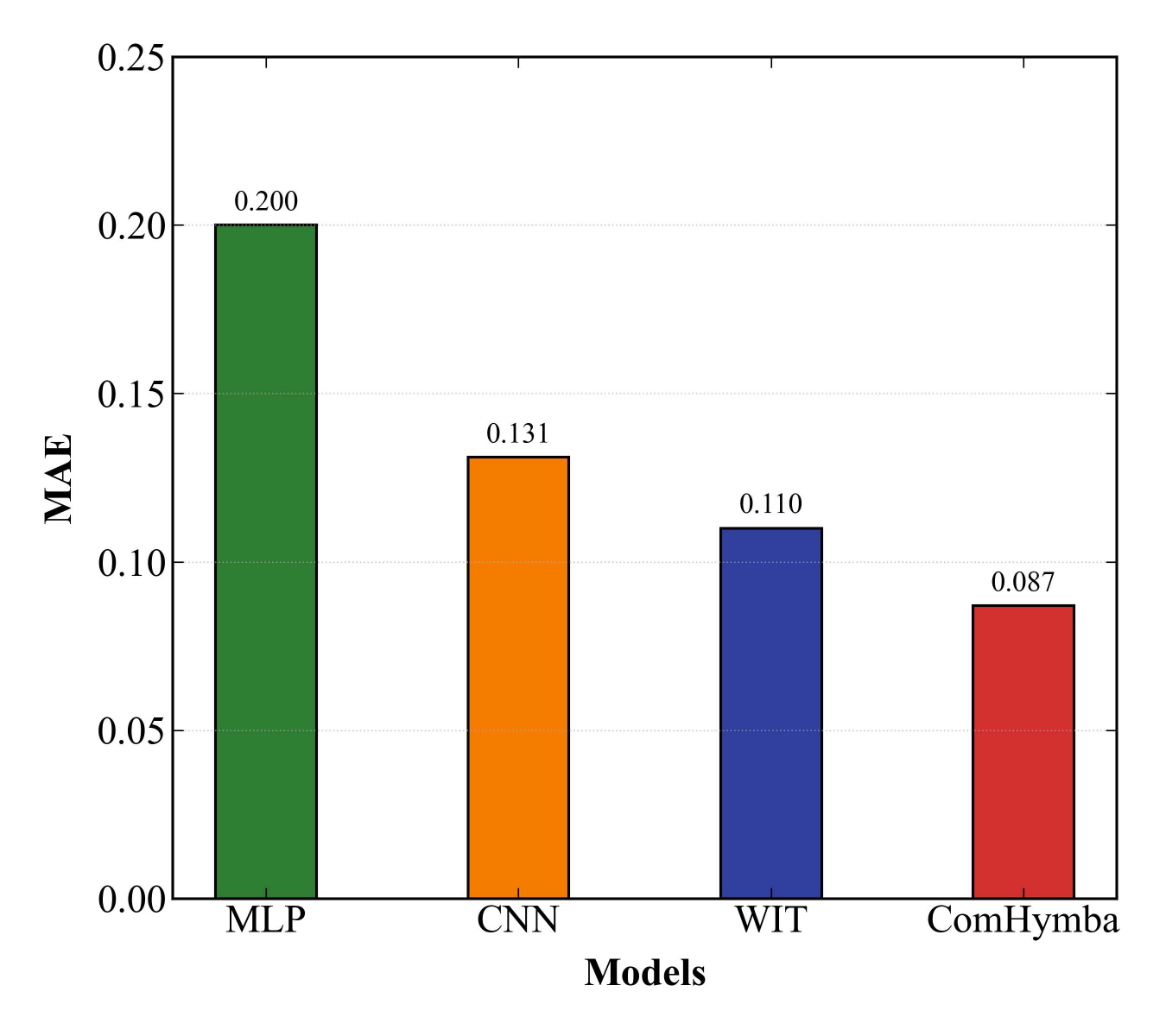}
        \caption{MAE for Path Loss Prediction.}
        \label{fig:pl_mae}
    \end{subfigure}
    
    \vspace{10pt}
    
    % 第二行
    \begin{subfigure}[b]{0.38\textwidth}
        \centering
        \includegraphics[width=\linewidth]{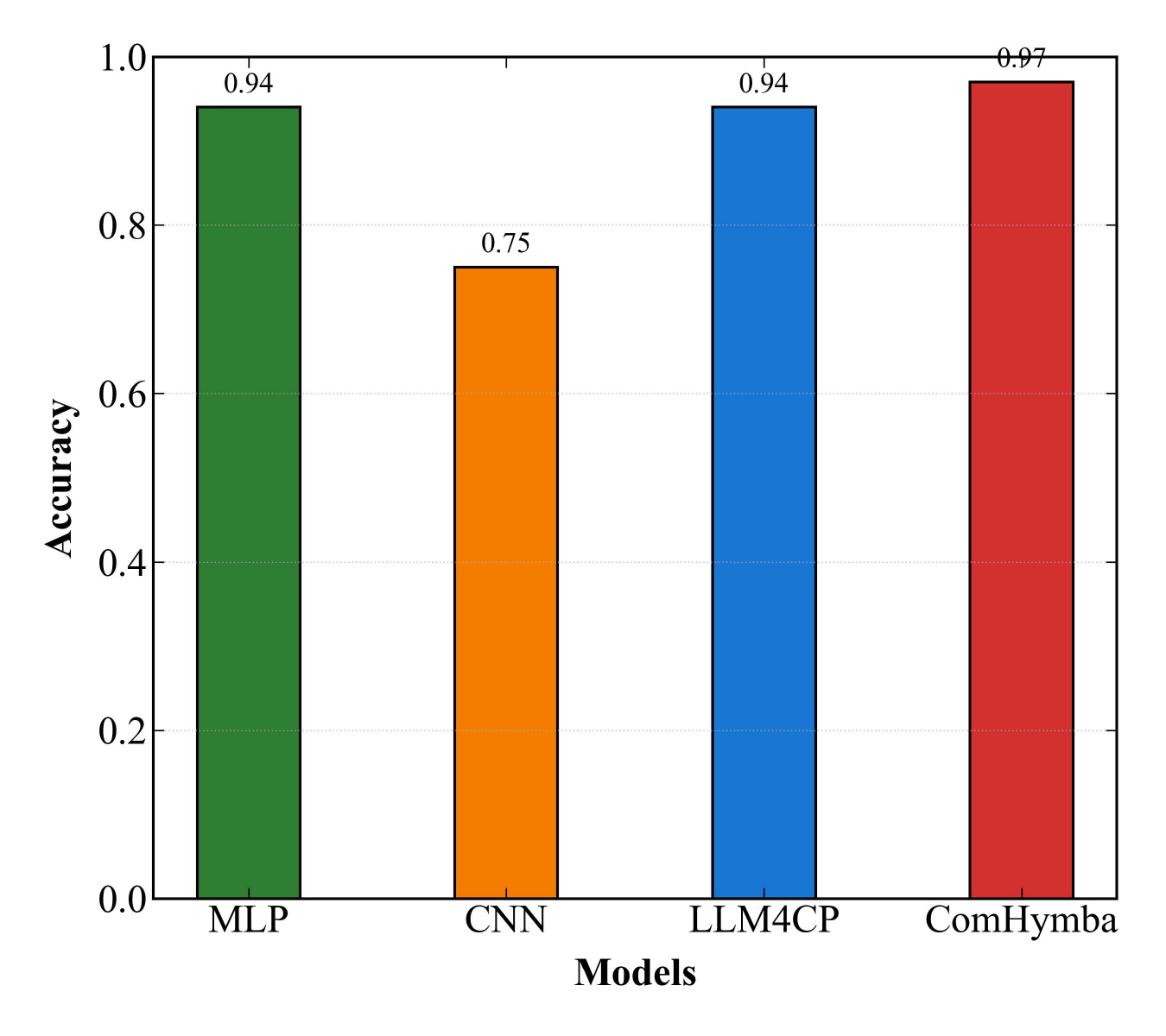}
        \caption{Classification accuracy across LoS/NLoS identification.}
        \label{fig:los_acc}
    \end{subfigure}
    \hspace{0.08\linewidth}
    \begin{subfigure}[b]{0.38\textwidth}
        \centering
        \includegraphics[width=\linewidth]{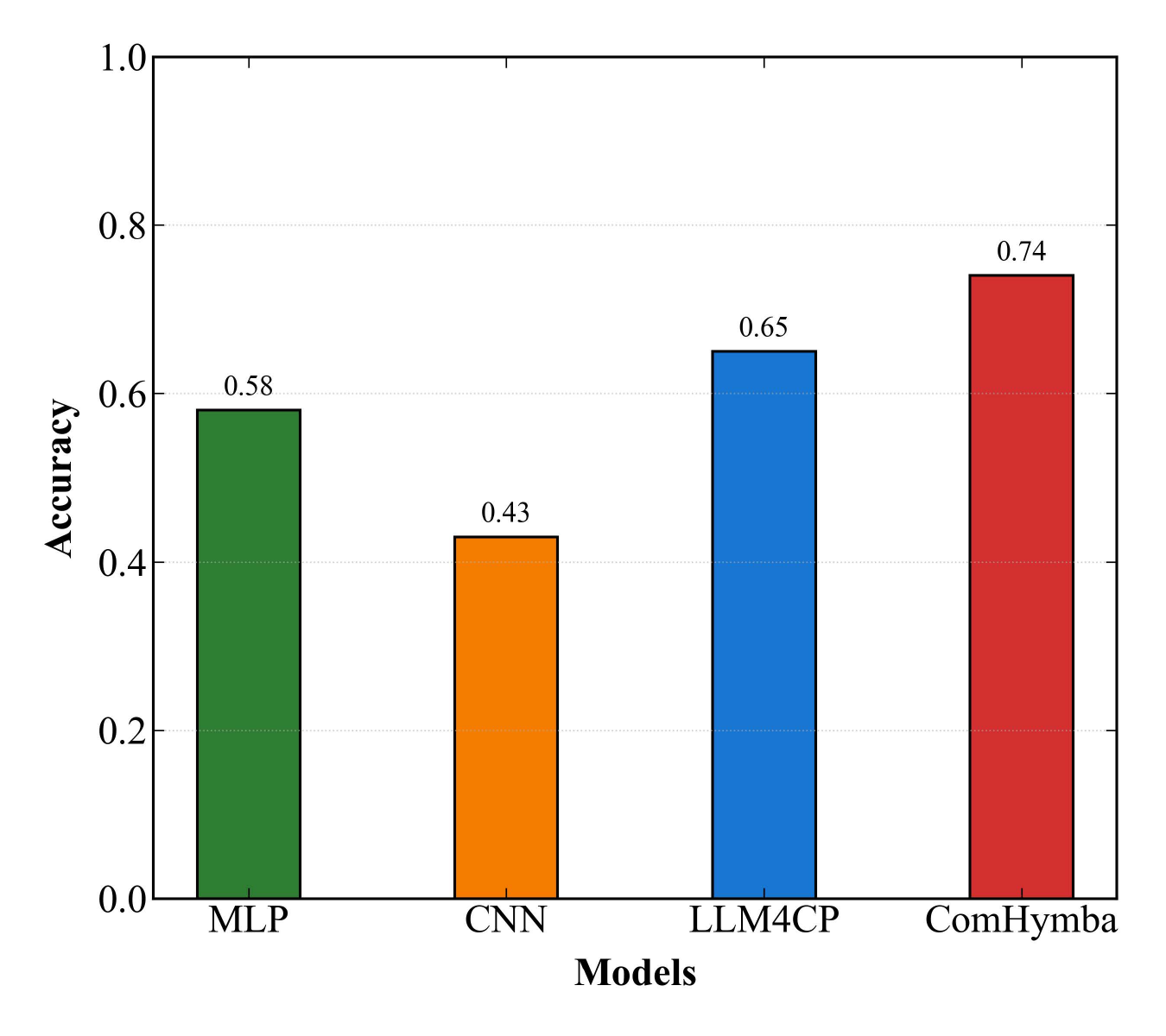}
        \caption{Classification accuracy across diverse wireless scenarios.}
        \label{fig:scenario_acc}
    \end{subfigure}
    
    \caption{Experimental results for environmental sensing tasks, including spatial regression and semantic classification.}
    \label{fig:group_sensing}
\end{figure*}

The environmental sensing tasks are designed to verify whether the ComHymba foundation model can extract macro-level spatial semantics from underlying physical signals. Based on this physical foundation, four specific sub-tasks are designed:
\begin{itemize}
    \item The user localization task maps micro-phase differences in the beam and frequency domains to macro-level Euclidean coordinates. This task verifies whether the features refined by ComHymba possess the high spatial resolution required for centimeter-level localization.

    \item The path loss prediction task requires the model to accurately predict the total power attenuation based on channel features. By leveraging ComHymba’s strong denoising capability, the model extracts stable large-scale attenuation trends from instantaneous channel data contaminated by severe fast-fading fluctuations, assisting the base station in optimizing real-time power allocation.

    \item The LoS/NLoS identification task tests the model's ability to distinguish different propagation modes and verifies if ComHymba's representations capture a deep understanding of propagation path integrity.

    \item The scenario identification task requires the model to categorize the specific wireless propagation environments, thereby verifying its capability to extract high-level semantic abstractions from physical channel responses.
\end{itemize}

Across sensing tasks, ComHymba demonstrates prominent advantages. In regression (Fig. \ref{fig:loc_mae}, Fig. \ref{fig:pl_mae}), it achieves a localization MAE of 0.054~m (a 46\% gain over WIT's 0.100~m) and a path loss MAE of 0.087~dB (outperforming WIT's 0.110~dB), validating its superior feature denoising and manifold alignment. In classification (Fig. \ref{fig:los_acc}, Fig. \ref{fig:scenario_acc}), ComHymba yields a 0.97 LoS identification accuracy (surpassing LLM4CP's 0.94 and CNN's 0.75) and tops scenario classification at 0.74 (versus LLM4CP's 0.65 and MLP's 0.58). These results confirm that ComHymba captures the implicit environmental scattering topology, rendering its universal representations a robust semantic descriptor.

\begin{figure*}[t]
    \centering
    \begin{subfigure}[b]{0.38\textwidth}
        \centering
        \includegraphics[width=\linewidth]{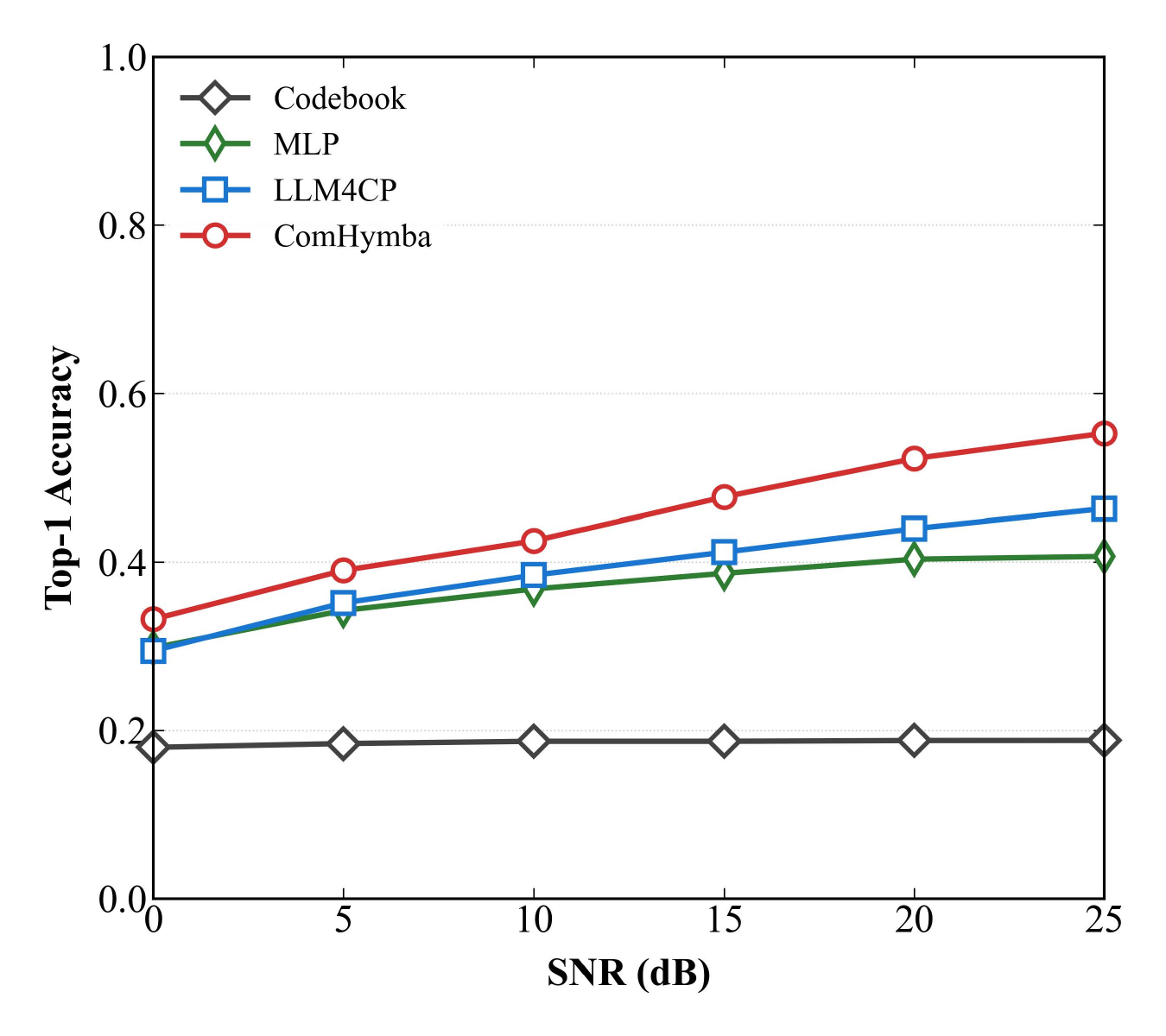}
        \caption{Top-1 beam selection accuracy versus SNR.}
        \label{fig:beam_top1}
    \end{subfigure}
    \hspace{0.08\linewidth}
    \begin{subfigure}[b]{0.38\textwidth}
        \centering
        \includegraphics[width=\linewidth]{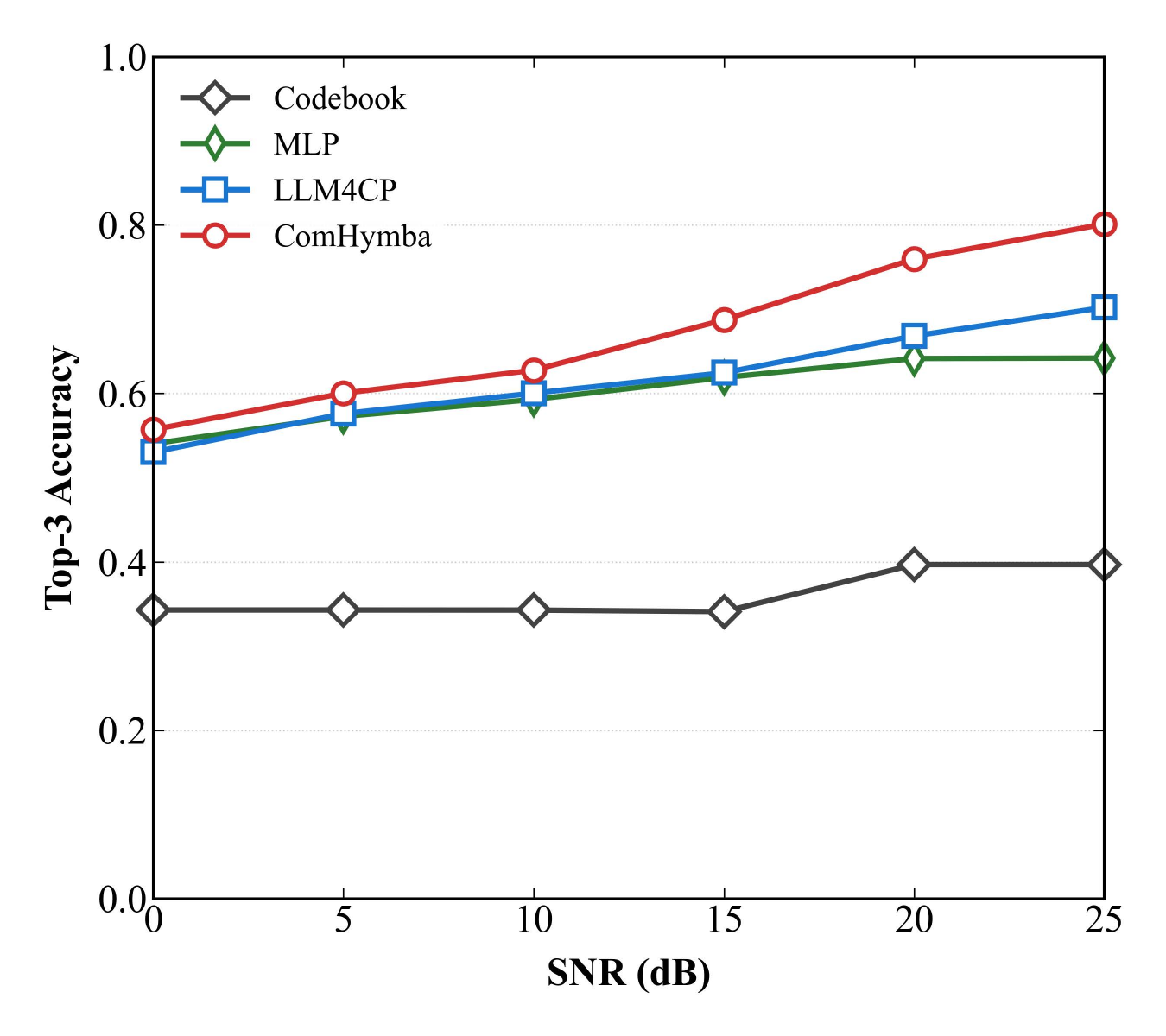}
        \caption{Top-3 beam selection accuracy versus SNR.}
        \label{fig:beam_top3}
    \end{subfigure}
    
    \caption{Comparison of beamforming decision accuracy across different SNR levels: (a) Top-1 and (b) Top-3.}
    \label{fig:group_beamforming}
\end{figure*}

\paragraph{Beam Management}

In millimeter-wave (mmWave) communications, traditional exhaustive beam sweeping schemes face significant challenges regarding signaling overhead and latency due to the extremely high dimensionality of the beam space and the susceptibility of channels to environmental blockages. Existing literature~\cite{stm} has pointed out the possibility of utilizing sub-6 GHz channel information to predict mmWave frequency beams. The physical foundation of this concept lies in spatial congruency; although fading characteristics vary across frequency bands, the dominant geometric propagation paths, such as the angle-of-arrival and angle-of-departure, exhibit strong cross-band correlation within the same physical environment. The ComHymba framework is built upon this physical mechanism, achieving precise prediction of optimal mmWave beams by extracting intrinsic spatial features from sub-6 GHz matrices.

Specifically, the beam management task requires retrieving the optimal beam weight vector $w_t$ from a predefined codebook $\mathcal{W} \in \mathbb{C}^{N_v \times N_c}$. To meet the requirements for high spatial resolution, this study employs a super-resolution discrete Fourier transform (DFT) codebook, resulting in a codebook size $N_c$ that is significantly larger than the number of physical antennas $N_v$. Through its robust feature refinement capabilities, the ComHymba model maps microscopic signal fluctuations in the sub-6 GHz band into high-density decision fingerprints for the mmWave band, thereby circumventing the burdensome downlink beam sweeping process.

Experimental results quantify and confirm the significant advantages of ComHymba in decision accuracy compared to existing state-of-the-art models. As shown in Fig. \ref{fig:beam_top1}, ComHymba demonstrates a clear lead in cross-band Top-1 prediction accuracy. At an SNR of 25 dB, its Top-1 accuracy reaches approximately 0.55, providing a substantial classification gain over LLM4CP and traditional MLP models. In the Top-3 accuracy evaluation, which offers a higher engineering reference value (as illustrated in Fig. \ref{fig:beam_top3}), the accuracy of ComHymba jumps to approximately 0.80 at 25 dB. This demonstrates that ComHymba, as a foundation model, can accurately extract frequency-independent underlying physical essences, ensuring the access reliability of high-frequency links while drastically reducing feedback overhead.

\subsubsection{Inference Latency and Scalability Analysis}

\begin{table}[!t]
\caption{Quantitative Comparison of Inference Latency (ms): ComHymba vs. Transformer}
\label{tab_latency_complete}
\centering
\resizebox{\columnwidth}{!}{
\begin{tabular}{ccccc}
\toprule
\makecell[c]{\textbf{Model Scale}} & \textbf{Input Dim} & \textbf{ComHymba} & \textbf{Transformer} & \textbf{Speedup} \\
\midrule
\multirow{4}{*}{\makecell{Small \\ ($\approx$100M)}} & (16, 32, 32) & 7.02    & 9.17     & 1.31$\times$ \\
                      & (32, 32, 32) & 13.11   & 19.93    & 1.52$\times$ \\
                      & (32, 32, 64) & 25.24   & 46.97    & 1.86$\times$ \\
                      & (32, 64, 64) & 47.32   & 154.61   & 3.27$\times$ \\
\midrule
\multirow{4}{*}{\makecell{Medium \\ ($\approx$200M)}} & (16, 32, 32) & 11.28   & 17.10    & 1.52$\times$ \\
                      & (32, 32, 32) & 20.74   & 36.45    & 1.76$\times$ \\
                      & (32, 32, 64) & 42.37   & 95.77    & 2.26$\times$ \\
                      & (32, 64, 64) & 94.86   & 310.93   & 3.28$\times$ \\
\midrule
\multirow{4}{*}{\makecell{Large \\ ($\approx$400M)}} & (16, 32, 32) & 21.84   & 32.75    & 1.50$\times$ \\
                      & (32, 32, 32) & 41.81   & 70.18    & 1.68$\times$ \\
                      & (32, 32, 64) & 88.86   & 183.21   & 2.06$\times$ \\
                      & (32, 64, 64) & 190.72  & 605.17   & 3.17$\times$ \\
\midrule
\multirow{4}{*}{\makecell{Huge \\ ($\approx$800M)}} & (16, 32, 32) & 42.35   & 68.50    & 1.62$\times$ \\
                      & (32, 32, 32) & 85.23   & 140.10   & 1.64$\times$ \\
                      & (32, 32, 64) & 175.21  & 360.51   & 2.06$\times$ \\
                      & (32, 64, 64) & \textbf{394.14} & \textbf{1300.20} & \textbf{3.30$\times$} \\
\bottomrule
\end{tabular}
}
\end{table}

Following the performance validation of downstream tasks, this section provides a quantitative analysis of the inference latency of the ComHymba architecture, emphasizing its scalability across varying model scales and input dimensions. The experimental evaluations are conducted on a high-performance computing platform utilizing an NVIDIA A40 (48GB VRAM) GPU. To evaluate the architectural robustness of our research framework, we tested model parameter scales ranging from 100M to 800M and four representative input tensor dimensions $(L, K, N_s)$ ranging from $(16, 32, 32)$ to $(32, 64, 64)$. 

The experimental results, summarized in Table \ref{tab_latency_complete}, demonstrate that ComHymba exhibits a significant computational advantage over the traditional pure Transformer architecture, with the performance gap widening significantly as the system scale increases. While the pure Transformer architecture is constrained by the quadratic complexity of self-attention mechanisms, leading to a ``computational wall'' as input dimensions grow, ComHymba utilizes a hybrid architecture with linear complexity to maintain efficient processing. As shown in Table \ref{tab_latency_complete}, this advantage is most prominent at the 800M parameter scale with the maximum $(32, 64, 64)$ input dimension, where the Transformer's latency escalates to $1300.20 \text{ ms}$, whereas ComHymba maintains a significantly lower latency of $394.14 \text{ ms}$. Furthermore, the speedup factor relative to the Transformer improves from approximately $1.31\times$ at smaller configurations to over $3.30\times$ at the 800M scale, confirming that the architectural benefits of ComHymba are amplified as both the parameter and input sizes expand. 

This scalability is critical for the future development of large-scale AI-native foundation models, which must handle the massive high-dimensional CSI data generated by extremely large antenna arrays. By effectively decoupling processing time from the quadratic growth of spatial and temporal dimensions, ComHymba provides a robust and scalable intelligence backbone; even as communications move toward ultra-high-resolution sensing and complex spatial manifolds, the architecture remains computationally viable. The ability to sustain low latency under increased parameter scales ensures that the model can capture deeper physical representations without sacrificing the throughput required for next-generation wireless systems.

\section{Conclusion}

In this paper, we have presented ComHymba, a novel domain-aware wireless foundation model designed to address the multifaceted challenges of AI-native networks. By integrating the asymmetric masked autoencoder framework with the hybrid-head Hymba architecture, ComHymba successfully bridges the gap between high-fidelity physical modeling and computational efficiency. The introduction of 3D spatio-temporal-frequency patchification, rotary positional embedding, and a decoupled amplitude-phase joint loss function ensures that the model captures the intrinsic electromagnetic properties of wireless channels while remaining robust to noise and data scarcity.

Our comprehensive evaluation across eight critical downstream tasks confirms that ComHymba consistently outperforms specialized state-of-the-art baselines in both accuracy and generalization. Furthermore, the inference latency analysis highlights ComHymba’s exceptional scalability, achieving up to a 3.3$\times$ speedup over traditional Transformer architectures at large scales. Ultimately, ComHymba serves as a unified and efficient intelligence backbone, providing a scalable and robust foundation for the realization of next-generation multi-functional communication and sensing systems.

\end{document}